\documentclass[ALICE,manyauthors]{cernphprep}
\usepackage[comma,square,numbers,sort&compress]{natbib}
\usepackage{hyperref}
\usepackage{xspace}
\usepackage[dvipsnames]{xcolor}
\usepackage{subfigure}
\usepackage[T1]{fontenc} % this allows us to type the ">" symbol without getting the weird upside down question mark!
\usepackage{bm}

\begin{document}
%%%%%%%%%%%%%%%%%%%%%%%%%%%%%%%%%%%%%%%%%%%%%%%%%%
% These are some new commands that may be useful 
% for paper writing in general. If other newcommands
% are needed for your specific paper, please feel 
% free to add here. 
%
% The currently available commands are organized in: 
% 1) Systems
% 2) Quantities
% 3) Energies and units
% 4) Detectors
% 5) particle species 
%%%%%%%%%%%%%%%%%%%%%%%%%%%%%%%%%%%%%%%%%%%%%%%%%%

% 0) our own custom commands
\newcommand{\redbf}[1]{\textcolor{red}{\textbf{#1}}}
\newcommand{\bluebf}[1]{\textcolor{blue}{\textbf{#1}}}
\newcommand{\greenbf}[1]{\textcolor{ForestGreen}{\textbf{#1}}}
\newcommand{\orangebf}[1]{\textcolor{YellowOrange}{\textbf{#1}}}
\newcommand{\arrow}{\ensuremath{\rightarrow}}

% 1) SYSTEMS 
\newcommand{\pp}           {pp\xspace}
\newcommand{\ppbar}        {\mbox{$\mathrm {p\overline{p}}$}\xspace}
\newcommand{\XeXe}         {\mbox{Xe--Xe}\xspace}
\newcommand{\PbPb}         {\mbox{Pb--Pb}\xspace}
\newcommand{\pA}           {\mbox{pA}\xspace}
\newcommand{\pPb}          {\mbox{p--Pb}\xspace}
\newcommand{\Pbp}          {\mbox{Pb--p}\xspace}
\newcommand{\AuAu}         {\mbox{Au--Au}\xspace}
\newcommand{\dAu}          {\mbox{d--Au}\xspace}

% 2) QUANTITIES 
\newcommand{\s}            {\ensuremath{\sqrt{s}}\xspace}
\newcommand{\snn}          {\ensuremath{\sqrt{s_{\mathrm{NN}}}}\xspace}
\newcommand{\pt}           {\ensuremath{p_{\rm T}}\xspace}
\newcommand{\meanpt}       {$\langle p_{\mathrm{T}}\rangle$\xspace}
\newcommand{\ycms}         {\ensuremath{y_{\rm cms}}\xspace}
\newcommand{\ylab}         {\ensuremath{y_{\rm lab}}\xspace}
\newcommand{\etarange}[1]  {\mbox{$\left | \eta \right |<#1$}}
\newcommand{\etaranges}[2] {\mbox{$#1<\eta <#2$}}
\newcommand{\yrange}[1]    {\mbox{$\left | y \right |<#1$}}
\newcommand{\dndy}         {\ensuremath{\mathrm{d}N_\mathrm{ch}/\mathrm{d}y}\xspace}
\newcommand{\dndeta}       {\ensuremath{\mathrm{d}N_\mathrm{ch}/\mathrm{d}\eta}\xspace}
\newcommand{\avdndeta}     {\ensuremath{\langle\dndeta\rangle}\xspace}
\newcommand{\dNdy}         {\ensuremath{\mathrm{d}N_\mathrm{ch}/\mathrm{d}y}\xspace}
\newcommand{\Npart}        {\ensuremath{N_\mathrm{part}}\xspace}
\newcommand{\Ncoll}        {\ensuremath{N_\mathrm{coll}}\xspace}
\newcommand{\dEdx}         {\ensuremath{\textrm{d}E/\textrm{d}x}\xspace}
\newcommand{\RpPb}         {\ensuremath{R_{\rm pPb}}\xspace}
\newcommand{\Raa}          {\ensuremath{R_{\rm AA}}\xspace}
\newcommand{\mmumu}        {\ensuremath{m_{\mu\mu}}\xspace} 

% 3) ENERGIES, UNITS
\newcommand{\nineH}        {$\sqrt{s}=0.9$~Te\kern-.1emV\xspace}
\newcommand{\seven}        {$\sqrt{s}=7$~Te\kern-.1emV\xspace}
\newcommand{\twoH}         {$\sqrt{s}=0.2$~Te\kern-.1emV\xspace}
\newcommand{\twosevensix}  {$\sqrt{s}=2.76$~Te\kern-.1emV\xspace}
\newcommand{\five}         {$\sqrt{s}=5.02$~Te\kern-.1emV\xspace}
\newcommand{\twosevensixnn}{$\sqrt{s_{\mathrm{NN}}}=2.76$~Te\kern-.1emV\xspace}
\newcommand{\fivenn}       {$\sqrt{s_{\mathrm{NN}}}=5.02$~Te\kern-.1emV\xspace}
\newcommand{\eightnn}      {$\sqrt{s_{\mathrm{NN}}}=8.16$~Te\kern-.1emV\xspace}
\newcommand{\eightnnBold}  {$\mathbf{\sqrt{\bm{s}_{\mathbf{NN}}}=8.16}$~Te\kern-.1emV\xspace}
\newcommand{\fivennBold}   {$\mathbf{\sqrt{\bm{s}_{\mathbf{NN}}}=5.02}$~Te\kern-.1emV\xspace}
\newcommand{\LT}           {L{\'e}vy-Tsallis\xspace}
\newcommand{\GeVc}         {Ge\kern-.1emV/$c$\xspace}
\newcommand{\MeVc}         {Me\kern-.1emV/$c$\xspace}
\newcommand{\TeV}          {Te\kern-.1emV\xspace}
\newcommand{\GeV}          {Ge\kern-.1emV\xspace}
\newcommand{\MeV}          {Me\kern-.1emV\xspace}
\newcommand{\GeVmass}      {Ge\kern-.1emV/$c^2$\xspace}
\newcommand{\MeVmass}      {Me\kern-.1emV/$c^2$\xspace}
\newcommand{\lumi}         {\ensuremath{\mathcal{L}}\xspace}

% 4) DETECTORS 
\newcommand{\ITS}          {\rm{ITS}\xspace}
\newcommand{\TOF}          {\rm{TOF}\xspace}
\newcommand{\ZDC}          {\rm{ZDC}\xspace}
\newcommand{\ZDCs}         {\rm{ZDCs}\xspace}
\newcommand{\ZNA}          {\rm{ZNA}\xspace}
\newcommand{\ZNC}          {\rm{ZNC}\xspace}
\newcommand{\SPD}          {\rm{SPD}\xspace}
\newcommand{\SDD}          {\rm{SDD}\xspace}
\newcommand{\SSD}          {\rm{SSD}\xspace}
\newcommand{\TPC}          {\rm{TPC}\xspace}
\newcommand{\TRD}          {\rm{TRD}\xspace}
\newcommand{\VZERO}        {\rm{V0}\xspace}
\newcommand{\VZEROA}       {\rm{V0A}\xspace}
\newcommand{\VZEROC}       {\rm{V0C}\xspace}
\newcommand{\Vdecay} 	   {\ensuremath{V^{0}}\xspace}

% 4) PARTICLE SPECIES 
\newcommand{\ee}           {\ensuremath{e^{+}e^{-}}}
\newcommand{\mumu}         {\ensuremath{\mu^+\mu^-}}
\newcommand{\pip}          {\ensuremath{\pi^{+}}\xspace}
\newcommand{\pim}          {\ensuremath{\pi^{-}}\xspace}
\newcommand{\kap}          {\ensuremath{\rm{K}^{+}}\xspace}
\newcommand{\kam}          {\ensuremath{\rm{K}^{-}}\xspace}
\newcommand{\pbar}         {\ensuremath{\rm\overline{p}}\xspace}
\newcommand{\kzero}        {\ensuremath{{\rm K}^{0}_{\rm{S}}}\xspace}
\newcommand{\lmb}          {\ensuremath{\Lambda}\xspace}
\newcommand{\almb}         {\ensuremath{\overline{\Lambda}}\xspace}
\newcommand{\Om}           {\ensuremath{\Omega^-}\xspace}
\newcommand{\Mo}           {\ensuremath{\overline{\Omega}^+}\xspace}
\newcommand{\X}            {\ensuremath{\Xi^-}\xspace}
\newcommand{\Ix}           {\ensuremath{\overline{\Xi}^+}\xspace}
\newcommand{\Xis}          {\ensuremath{\Xi^{\pm}}\xspace}
\newcommand{\Oms}          {\ensuremath{\Omega^{\pm}}\xspace}
\newcommand{\degree}       {\ensuremath{^{\rm o}}\xspace}

%%%%%%%%%%%%%%  Title page %%%%%%%%%%%%%%%%%%%%%%%%
\begin{titlepage}
% the dates below correspond to CERN approval
% please don't touch: EB chairs will take care
\PHyear{2020}       % required, will be obtained from CERN
\PHnumber{090}      % required, will be obtained from CERN
\PHdate{20 May}  % required, will be obtained from CERN
%%%%%%%%%%%%%%%%%%%%%%%%%%%%%%%%%%%%%%%%%%%%%%%%%%%%

%%% Put your own title + short title here:
\title{Z-boson production in \pPb collisions at \eightnnBold \\ and \PbPb collisions at \fivennBold} \ShortTitle{Z-boson production in p--Pb (Pb--Pb) collisions at $\sqrt{s_{\mathrm{NN}}}$ = 8.16 (5.02) TeV}   % appears on left page headers

%%% Do not change the next lines
\Collaboration{ALICE Collaboration\thanks{See Appendix~\ref{app:collab} for the list of collaboration members}}
\ShortAuthor{ALICE Collaboration} % appears on right page headers, do not change

\begin{abstract}
Measurement of Z-boson production in \pPb collisions at \eightnn and \PbPb collisions at \fivenn is reported. It is performed in the dimuon decay channel, through the detection of muons with pseudorapidity $-4 < \eta_{\mu} < -2.5$ and transverse momentum $\pt^{\mu} > 20$ \GeVc in the laboratory frame. The invariant yield and nuclear modification factor are measured for opposite-sign dimuons with invariant mass $60 < \mmumu < 120$ \GeVmass and rapidity $2.5 < \ycms^{\mu\mu} < 4$. They are presented as a function of rapidity and, for the \PbPb collisions, of centrality as well. The results are compared with theoretical calculations, both with and without nuclear modifications to the Parton Distribution Functions (PDFs). In \pPb collisions the center-of-mass frame is boosted with respect to the laboratory frame, and the measurements cover the backward ($-4.46<\ycms^{\mu\mu}<-2.96$) and forward ($2.03<\ycms^{\mu\mu}<3.53$) rapidity regions. For the \pPb collisions, the results are consistent within experimental and theoretical uncertainties with calculations that include both free-nucleon and nuclear-modified PDFs. For the \PbPb collisions, a $3.4\sigma$ deviation is seen in the integrated yield between the data and calculations based on the free-nucleon PDFs, while good agreement is found once nuclear modifications are considered.
\end{abstract}
\end{titlepage}

\setcounter{page}{2} %please do not remove this line

%%%%%%%%%%%%%%%%%%%%%%%%%%%%%%%%
% begin main text
%%%%%%%%%%%%%%%%%%%%%%%%%%%%%%%%

\section{Introduction}

Measurements of W and Z electroweak vector boson production are useful probes for studying the initial conditions of heavy-ion collisions. Their production occurs predominantly via the Drell-Yan process, in which a quark-antiquark pair annihilates into a lepton pair~\cite{PhysRevLett.25.316,H.:2020ecd}. At leading order, this is an electroweak process although at higher orders gluon radiation must be accounted for. Due to the large masses of these resonances, vector boson production occurs in the early stages of the collisions and their cross section in elementary parton--parton interactions can be calculated within perturbative Quantum Chromodynamics (pQCD). The large masses also allow for high-precision theoretical calculations, currently reaching up to Next-to-Next-to-Leading Order (NNLO) accuracy~\cite{Catani:2009sm,Anastasiou:2003ds}.

Since neither the vector bosons nor their leptonic decay products carry color charge, they do not interact strongly with the dense QCD medium formed in heavy-ion collisions.

There are hints that the muons undergo electromagnetic interactions with the dense QCD medium, seen by \pt broadening of dimuon spectra~\cite{Aaboud:2018eph}. However, this broadening can also be described by photo-production~\cite{Zha:2018tlq}. Regardless of the physical origin, the scale of these \pt-modifications is negligible compared to the average momentum of the muons, which is about half of the Z-boson mass. Thus, the information carried by the muons is not diluted due to final state interactions, allowing to probe the initial state directly.
At the Large Hadron Collider (LHC), the center-of-mass energies and luminosities are large enough to allow the production of these bosons to be measured in heavy-ion collisions.

Since the production of electroweak bosons occurs predominantly through quark-antiquark annihilation, it is dependent on the longitudinal momentum distributions of the quarks in the initial state of the collision. These distributions are parametrized by the Parton Distribution Functions (PDFs) $f_{\mathrm{i}} (x, Q^2)$~\cite{Paukkunen:2009ks}. Here, $f_{\mathrm{i}}$ gives the probability of finding a parton of type $\mathrm{i}$ (this could be either a gluon or a quark with a given flavor) with momentum fraction $x$ of the parent nucleon (also known as Bjorken-$x$) and squared 4-momentum transfer vector $Q^2$. In general, PDFs are obtained through global fits to data, combining information from multiple experiments. Most of these data come from Deep Inelastic Scattering (DIS) experiments, although data from the Tevatron and the LHC have recently been included as well~\cite{Butterworth:2015oua,Martin:2009iq,Martin:1999ww,Harland-Lang:2014zoa}.

It has been observed that in a nucleus with mass number $A$, the distributions of partons $f_{\mathrm{i}}^{\mathrm{A}} (x, Q^2)$ differ from the free-nucleon PDFs scaled by the number of protons and neutrons $A f_{\mathrm{i}}^{\mathrm{nucleon}} (x, Q^2)$~\cite{Aubert:1983xm}. The modified distributions can be described by means of so-called nuclear Parton Distribution Functions (nPDFs). The Bjorken-$x$ domain can be divided into four main regions, displaying various nuclear effects~\cite{Paukkunen:2010qg,Armesto:2006ph}. It should be noted that the precise values of the boundaries of the $x$-regions depend on the parton flavor, nPDF parametrization and $Q^{2}$. The following values assume $Q^{2} = M_{Z}^{2}$~\cite{Paukkunen:2010qg}. At low $x$, up to $x \sim 0.05$, a depletion of partons is present in nPDFs compared to free-nucleon PDFs. This depletion is referred to as shadowing. Then, in $x \sim 0.05 - 0.3$ an enhancement is seen, called antishadowing. Following this, another depletion region, the so-called EMC region, is present from $x \sim 0.3$ to $x \sim 0.9$. Lastly, $x{>}\sim 0.9$ to unity is the so-called Fermi region, where again an enhancement is present. These nuclear modifications to the PDFs influence the production of electroweak bosons~\cite{Vogt:2000hp}, but suffer from large uncertainties.
Since the parametrizations are obtained through global fits to data, experimental uncertainties enter the nPDFs as well.
Accurate measurements of W and Z bosons at the LHC can therefore help to constrain the nPDFs~\cite{Paukkunen:2010qg,Salgado:2011wc,Salgado:2011pf}, which are fundamental ingredients to properly describe the initial state of heavy-ion collisions. An in-depth overview of nPDFs can be found in Ref.~\cite{Paukkunen:2018kmm}.

The production of electroweak bosons at the LHC has already been studied in several collision systems, at various energies and rapidities~\cite{Alice:2016wka,Acharya:2017wpf,Aad:2015gta,Aad:2019lan,Aaboud:2018nic,Aad:2012ew,Aad:2014bha,Khachatryan:2015pzs,Chatrchyan:2011ua,Chatrchyan:2012nt,Chatrchyan:2014csa,Khachatryan:2015hha,Sirunyan:2019dox,Aaij:2014pvu}. 
At midrapidity, the data in different collision systems are generally well described by theoretical calculations both with and without nuclear modification of the PDFs ~\cite{Aad:2015gta,Aad:2019lan,Aaboud:2018nic,Aad:2012ew,Aad:2014bha,Khachatryan:2015pzs,Chatrchyan:2011ua,Chatrchyan:2012nt,Chatrchyan:2014csa,Khachatryan:2015hha,Sirunyan:2019dox}.
In fact, the covered Bjorken-$x$ range at midrapidity extends over the antishadowing and shadowing region (depending on the transition value, even into the EMC region). As a result, their competing effects reduce the final effect of nuclear modifications.
However, at larger rapidities and therefore lower $x$, there are increasingly stronger deviations between calculations with models that either do or do not include nuclear modification of the PDFs~\cite{Acharya:2017wpf, Sirunyan:2019dox}. 
The data taken at the LHC are in a kinematic regime in $(x, Q^2)$ which is also sensitive to contributions coming from quark flavors such as charm and strange, present as sea quarks in the nucleons~\cite{Kusina:2016fxy}. 
The uncertainties in both the nPDFs and the free-nucleon PDFs (where nuclear fixed-target data were also used for the fits) for these flavors are large and a combination of heavy-ion and proton data can help in reducing them~\cite{Kusina:2016fxy}.

This paper presents the measurement of the Z-boson production at forward rapidity through the dimuon decay channel in \pPb collisions at \eightnn as well as in \PbPb collisions at \fivenn. 
The \pPb measurement is the first at this energy, following an earlier ALICE publication with data taken at \fivenn~\cite{Alice:2016wka}. The Z-boson production in \PbPb collisions at \fivenn has already been published by the ALICE Collaboration using data collected in 2015~\cite{Acharya:2017wpf}. In 2018, new \PbPb data were collected at the same collision energy, corresponding to an integrated luminosity twice that of 2015. The dataset used in this paper includes both the 2015 and 2018 samples and therefore supersedes the previous \PbPb results. The larger dataset allows for a more differential analysis as well as increased precision on the integrated cross section measurement.

The paper is organized as follows: the ALICE detector and data samples are detailed in Sec.~\ref{sec:ALICEdetector}, followed by the analysis procedure in Sec.~\ref{sec:analysismethod}. The main results are then given in Sec.~\ref{sec:results} and the conclusions are drawn in Sec.~\ref{sec:conclusion}.

\section{ALICE detector and data samples}
\label{sec:ALICEdetector}
Z bosons are reconstructed via their dimuon decay channel using data from the ALICE muon spectrometer, which selects, identifies and reconstructs muons in the pseudorapidity range  $-4< \eta_{\mu} <-2.5$~\cite{ALICE:1999aa}. The tracking system consists of five stations, each containing two multi-wire proportional cathode pad chambers. The third station is located inside a dipole magnet that provides an integrated magnetic field of 3 $\mathrm{T} \times \mathrm{m}$. A conical absorber of 10 interaction lengths ($\lambda_{\mathrm{i}}$) 
made of carbon, concrete and steel, is located in front of the tracking system to filter out the hadrons and low-momentum muons from the decay of light particles (such as pions or kaons). 
The muon trigger system consists of four resistive plate chamber planes arranged in two stations placed downstream of an iron wall of $\sim$7.2~$\lambda_{\mathrm{i}}$ that reduces the contamination of residual hadrons leaking from the front absorber. 
Finally, a small-angle beam shield made of dense materials protects the whole spectrometer from secondary particles coming from beam-gas interactions and from interactions of large rapidity particles with the beam pipe.

Primary vertex reconstruction is performed by the Silicon Pixel Detector (\SPD), the two innermost cylindrical layer of the Inner Tracking System (\ITS)~\cite{Aamodt:2010aa}. 
The first and second layer cover the pseudorapidity regions $\etarange{2.0}$ and $\etarange{1.4}$, respectively.
Two arrays of scintillator counters (\VZEROA and \VZEROC~\cite{Abbas:2013taa}) are used to trigger events and to reject events from beam-gas interactions. The \VZEROA and \VZEROC detectors are located on both sides of the interaction point at $z=3.4$ m and $z=-0.9$ m 
and cover the pseudorapidity regions $2.8< \eta <5.1$ and $-3.7< \eta <-1.7$, respectively. The \VZERO detectors are also used to estimate the centrality in \PbPb collisions by using a Glauber model fit to the sum of their signal amplitudes~\cite{ALICE-PUBLIC-2018-011}. The events are then distributed in classes corresponding to a percentile of the total inelastic hadronic cross section.
Finally, the Zero Degree Calorimeters (\ZDC)~\cite{ALICE:2012aa}, placed on both sides of the interaction point at $z=\pm 112.5$ m, are used to reject electromagnetic background. A complete description of the ALICE detector and its performance can be found in Refs.~\cite{Aamodt:2008zz,Abelev:2014ffa}.

The analysis in \pPb collisions is performed on data collected in 2016 at a center-of-mass energy \eightnn. The data were taken in two beam configurations, with either the proton (p-going) or lead ion (Pb-going) moving towards the muon spectrometer. By convention, the proton moves toward positive rapidities. 
Because of the asymmetry in the proton and lead beam energies ($E_{\rm p}=6.5$ \TeV and $E_{\rm Pb}=2.56$ \TeV per nucleon), the resulting nucleon--nucleon center-of-mass system is boosted with respect to the laboratory frame by $\Delta \ycms = \pm 0.465$. Therefore, the rapidity acceptance of the muon spectrometer in the center-of-mass system is $2.03<\ycms^{\mu\mu}<3.53$ for the p-going configuration and $-4.46<\ycms^{\mu\mu}<-2.96$ for the Pb-going configuration.
The data used in the \PbPb analysis were taken in 2015 and 2018 at \fivenn and cover the rapidity\footnote{In the ALICE reference frame the muon spectrometer covers negative $\eta$. However, due to the symmetric nature of the \PbPb collisions, we use positive values for the probed rapidity interval.} interval $2.5<\ycms^{\mu\mu}<4$.

The events selected for the analyses require two opposite-sign muon candidates in the muon trigger system, each with a transverse momentum above a configurable threshold, in coincidence with a minimum bias (MB) trigger. 
The latter was defined by the coincidence of signals in the two arrays of the \VZERO detector. 
In the \PbPb analysis, only the events corresponding to the most central 90\% of the total inelastic cross section (0--90\%) are used. For these events the MB trigger is fully efficient and the contamination by electromagnetic interactions is negligible. 
For \pPb collisions, the Z-boson cross section is calculated using a luminosity normalization factor obtained via a reference process corresponding to the MB trigger condition itself. Therefore the MB trigger efficiency does not affect the cross section evaluation.
Finally, the muon trigger threshold was $\pt^{\mu} \gtrsim 0.5$ \GeVc for \pPb and $\pt^{\mu} \gtrsim 1$ \GeVc for \PbPb collisions.
After the event selection, the integrated luminosity in \PbPb collisions is about $750~\mu \rm{b}^{-1}$. In the \pPb analysis, where a precise value of the luminosity is needed to compute the Z-boson cross section, dedicated Van der Meer scans were performed~\cite{ALICE-PUBLIC-2018-002}. 
The values of the luminosity amount to $8.40 \pm 0.16~\rm{nb}^{-1}$ and $12.74 \pm 0.24~\rm{nb}^{-1}$ for the p-going and Pb-going configuration, respectively.

\section{Analysis procedure}
\label{sec:analysismethod}

The Z-boson signal extraction is performed by combining muons of high transverse momentum in pairs with opposite charge. Muon track candidates are reconstructed in the tracking system of the spectrometer using the algorithm described in Ref.~\cite{Aamodt:2011gj}. In order to ensure a clean data sample, a selection is performed on the single muon tracks reconstructed in the muon spectrometer, requiring them to have a pseudorapidity $-4 < \eta_{\mu} < -2.5$ and a polar angle measured at the end of the front absorber of $170\degree < \theta_{\rm abs} < 178\degree$. This procedure removes tracks at the edge of the spectrometer acceptance, and rejects tracks crossing the high-density section of the absorber, which experience significant multiple scattering. The background from tracks not pointing to the nominal interaction vertex, mostly coming from beam--gas interactions and muons produced in the front absorber, is removed by applying a selection on the product of the track momentum and its distance of closest approach to the primary vertex (i.e. the distance to the primary vertex of the track trajectory projected in the plane transverse to the beam axis). Finally, a track is identified as a muon if the track reconstructed in the tracking system matches a track segment in the triggering stations.

\begin{figure}[h]
  \centering
  \subfigure[]{\includegraphics[width=0.49\linewidth]{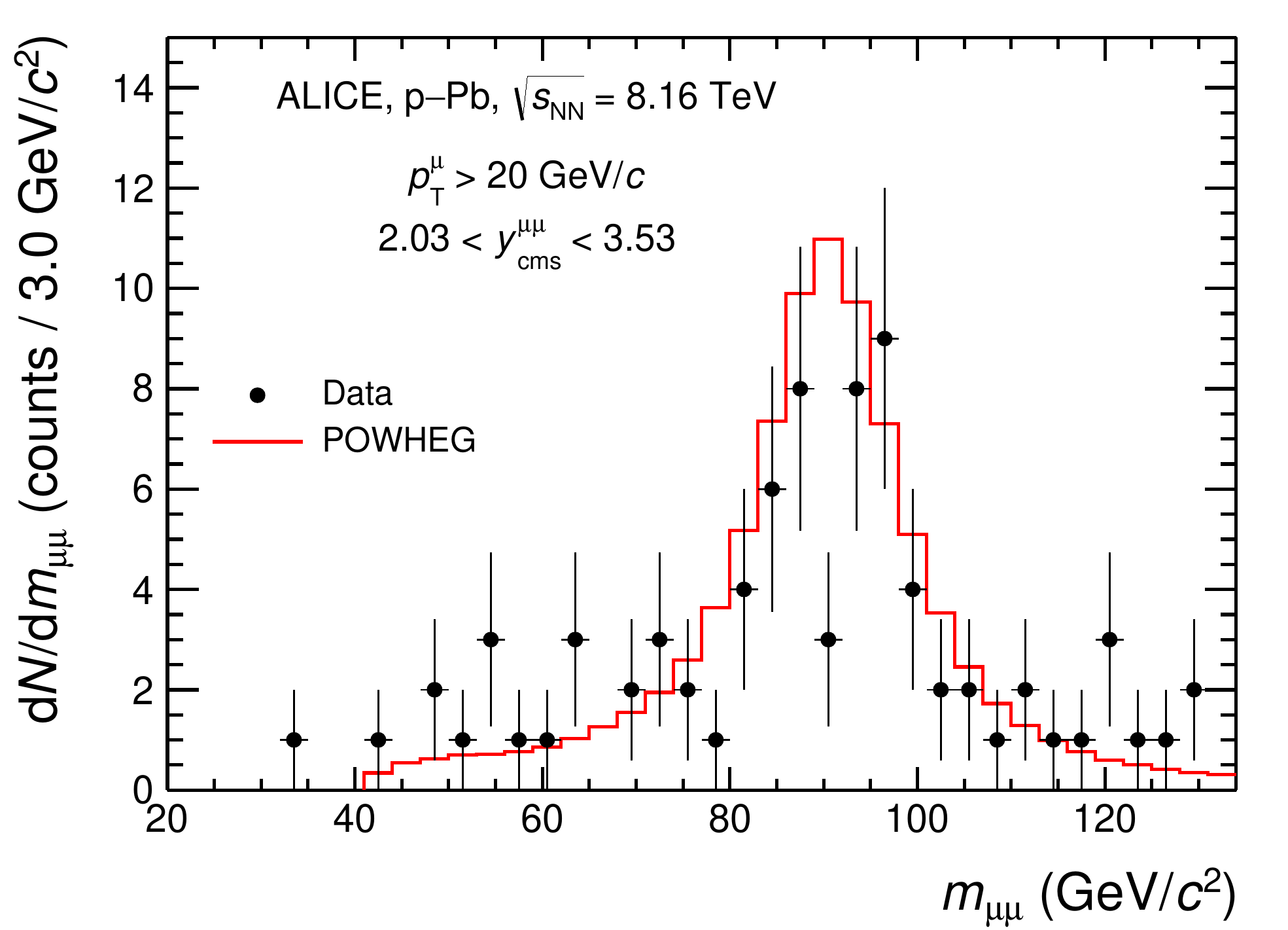}}
  \subfigure[]{\includegraphics[width=0.49\linewidth]{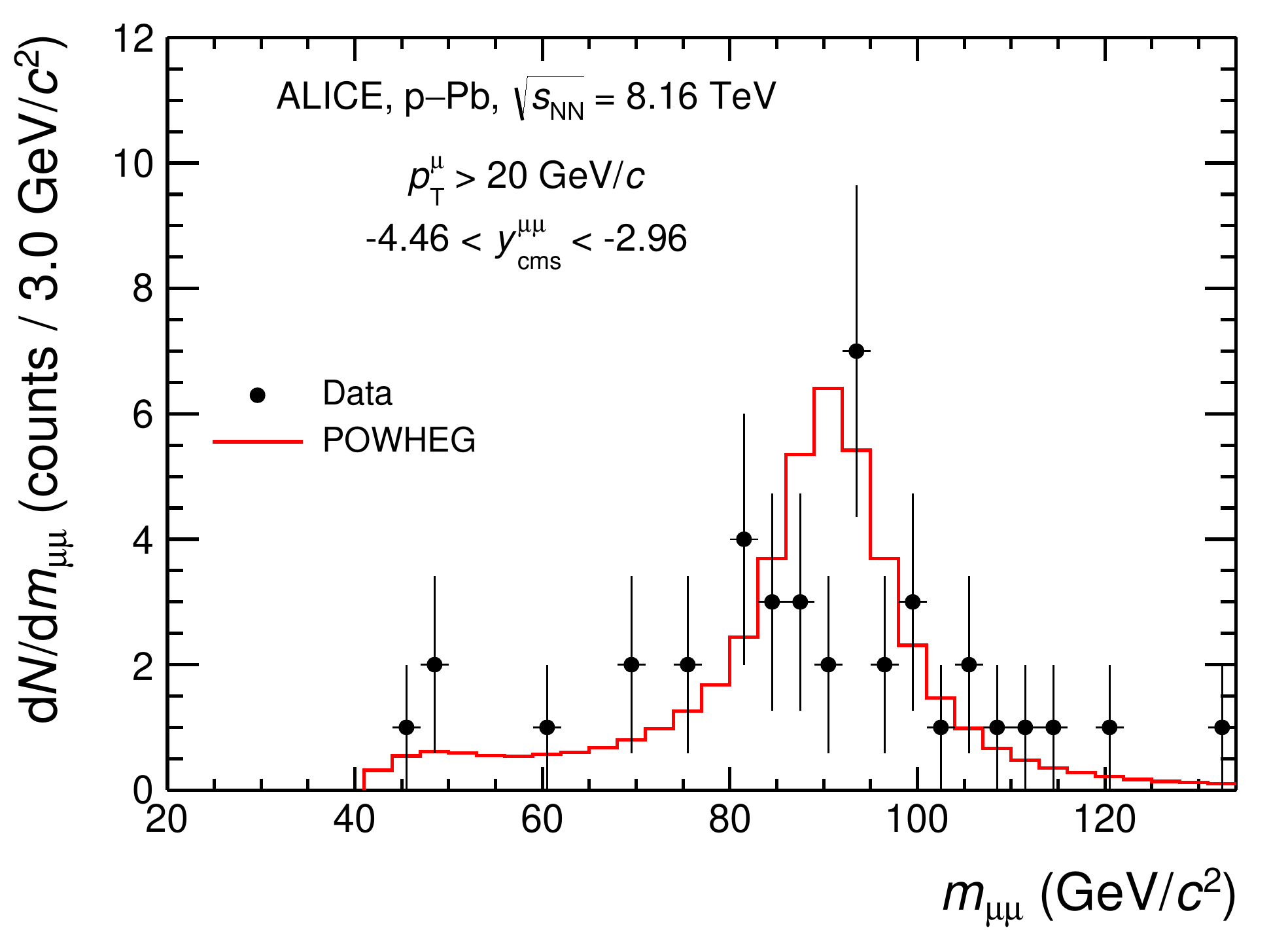}}
  \subfigure[]{\includegraphics[width=0.49\linewidth]{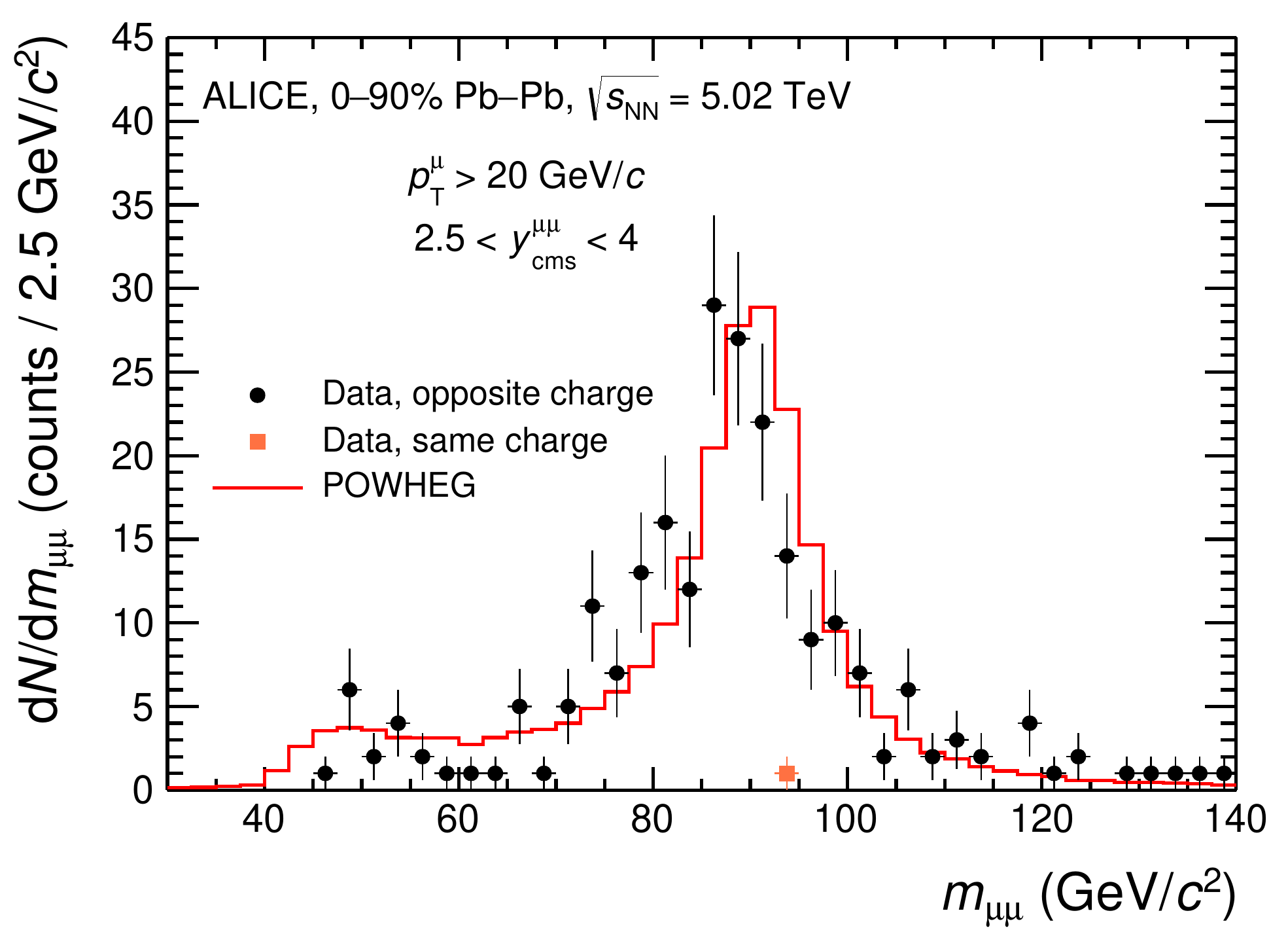}}
  \caption{Invariant mass distribution of $\mu^+ \mu^-$ pairs for p--Pb collisions at \eightnn in (a) the p-going and (b) Pb-going data samples, and (c) Pb--Pb collisions at \fivenn. The distributions are obtained from muons with $-4 < \eta_{\mu} < -2.5$ and $\pt^{\mu} > 20$ GeV/$c$ (black points) and compared with POWHEG simulations (red curves), which are normalized to the number of Z bosons in the data. The single like-sign dimuon entry is also shown (orange point) for Pb--Pb collisions, while no entries were found in the p--Pb samples.}
  \label{fig:Zcandidates}
\end{figure}

Only muons with $\pt^{\mu} > 20$ GeV/$c$ are used, to reduce the contribution from low-mass resonances and semileptonic decays of charm and beauty hadrons. The $\mu^+\mu^-$ pairs are counted in the invariant mass range $60 < \mmumu < 120$ GeV/$c^2$, where the Z-boson contribution is dominant with respect to the Drell-Yan process. The invariant mass distributions of the Z-boson candidates are shown in Fig.~\ref{fig:Zcandidates} for minimum bias p--Pb collisions in the p-going and Pb-going configurations, and Pb--Pb collisions in the centrality range 0--90\%. Several background sources can contribute to the invariant mass distributions of opposite-charge dimuons. The combinatorial background from random pairing of muons in an event is evaluated by looking at the like-sign pairs ($\mu^\pm\mu^\pm$), applying the same selection criteria as for the signal extraction. In the Pb--Pb sample, one pair is found in the invariant mass range considered, which is subtracted from the signal distribution. In p--Pb collisions, no such pairs are found in the region of interest. An upper limit for this background contribution is evaluated by releasing the $\pt^{\mu}$ selection, fitting the resulting invariant mass distribution between 2 and 50 GeV/$c^2$ and extrapolating the fit to the 60--120 GeV/$c^2$ range. Various functions of exponential and power law forms were tried. With this procedure, the number of same-charge events in the region of interest is much smaller than 1\% of the opposite-charge one, and is therefore neglected.

Contributions from ${\rm c\overline{c}}$, ${\rm b\overline{b}}$, ${\rm t\overline{t}}$ and the muon decay of $\tau$ pairs in the process Z $\rightarrow \tau^+\tau^- \rightarrow \mu^+\mu^- + X$ were estimated with Monte Carlo (MC) simulations using the POWHEG event generator~\cite{powheg}. In p--Pb collisions, the sum of these contributions amounts to 1\% of the signal in the p-going configuration, which is taken as a systematic uncertainty from this background source. This contribution is negligible for the Pb-going configuration. In Pb--Pb collisions, a value of 1\% is estimated as described in the previous publication~\cite{Acharya:2017wpf}.

The low amount of background allows the signal to be extracted by counting the candidates in the interval $60 < \mmumu < 120$ GeV/$c^2$ in the distributions shown in Fig.~\ref{fig:Zcandidates}. In the p--Pb data sample, 64 $\pm$ 8 (34 $\pm$ 6) good $\mu^+ \mu^-$ pairs are counted in the forward (backward) rapidity region. In Pb--Pb collisions, 208 $\pm$ 14 Z bosons are counted after merging the 2015 and 2018 data samples. All quoted uncertainties are statistical.

The dimuon invariant mass distributions are compared with the mass shapes obtained by detector-level simulations of the Z $\rightarrow \mu^+\mu^-$ process, generated using the POWHEG generator~\cite{powheg} paired with PYTHIA 6.425~\cite{pythia} for the parton shower. The CT10~\cite{ct10} free-nucleon PDFs are used, with EPS09NLO~\cite{eps09} for nuclear modifications. The propagation of the particles through the detector is simulated with the GEANT3 transport code~\cite{geant3}. To account for the modification of the production due to the light-quark flavor content of the nucleus (isospin effect), the simulated distributions are obtained with a weighted average of all possible binary collisions: proton--proton, proton--neutron and for Pb--Pb collisions also neutron--neutron. At high $\pt^{\mu}$, tracks are nearly straight so a small misalignment of the detector elements will generate large changes in the track parameters. Therefore, a detailed study of the alignment of the tracking chambers is of utmost importance in order to correctly reproduce the track reconstruction in the simulations. The absolute position of the detector elements was measured by photogrammetry before data taking. The relative position of the elements is then estimated using the MILLEPEDE~\cite{millipede} package, combining data taken with and without magnetic field, with a precision of about 100 $\mu$m. This residual misalignment is then taken into account in the simulations of the Z production and the efficiency computation. This method accounts for the relative misalignment of the detector elements but it is not sensitive to a global displacement of the entire muon spectrometer. The simulation of the response of the muon tracking system is based on a data-driven parametrization of the measured resolution of the clusters associated to a track~\cite{Abelev:2014ffa}, using extended Crystal-Ball (CB) functions~\cite{Gaiser:144456} to reproduce the distribution of the difference between the cluster and the track positions in each chamber. The CB functions, having a Gaussian core and two power law tails, are first tuned to data and then used to simulate the smearing of the track parameters. The effect of a global misalignment of the spectrometer is implemented by applying a systematic shift, in opposite directions for positive and negative tracks, to the distribution of the angular deviation of the tracks in the magnetic field. This shift is tuned to reproduce the observed difference in the $\pt^{\mu}$ distributions of positive and negative tracks. In Pb--Pb collisions, the data were taken with two opposite magnetic field polarities of the muon spectrometer dipole magnet. In this case, the sign of the shift is inverted accordingly.

The Z-boson raw yields are corrected for the acceptance times efficiency ($A \times \epsilon$) of the detector. It is evaluated with the MC simulations of the Z $\rightarrow \mu^+\mu^-$ process with POWHEG described above. The $A \times \epsilon$ is estimated as the ratio of reconstructed Z bosons with the same selections as for the data, to the number of generated ones with $2.5 < y_{\mathrm{lab}}^{\mu\mu} < 4$ for the dimuon pairs, and $\pt^{\mu} > 20$ GeV/$c$ and $-4 < \eta_{\mu} < -2.5$ for the muons. The dimuon invariant mass selection $60 < \mmumu < 120$ GeV/$c^2$ is applied to both reconstructed and generated distributions. In p--Pb collisions, the efficiency is 74\% (72\%) for the p-going (Pb-going) sample. In Pb--Pb collisions, the efficiency depends on the detector occupancy and therefore on the centrality of the collision. To account for this effect, the generated signal is embedded in real Pb--Pb events. The efficiency is found to be stable from peripheral to semi-central collisions, with a value of about 77\% (71\%) in the 2015 (2018) data sample and decreases to 71\% (66\%) for the most central collisions. The centrality-integrated efficiency amounts to 73\% in the 2015 dataset and 68\% for the 2018 dataset. The Z-boson invariant yield is then computed by dividing the number of measured candidates, corrected for $A \times \epsilon$, by the corresponding number of minimum bias events. The latter is evaluated using the normalization factor $F_{\mu{\rm -trig/MB}}$, corresponding to the inverse of the probability to observe an opposite-sign dimuon triggered event in a MB event. The value of $F_{\mu{\rm -trig/MB}}$ is evaluated with two methods: (i) by applying the opposite-sign dimuon trigger condition in the analysis of MB events, and (ii) by comparing the counting rate of the two triggers, both corrected for pile-up effects. The first method is performed on the smaller data sample of the recorded MB events. In the second method, information from the trigger counters was used. This means that the relative frequencies of MB and triggered events were counted, including events that were not stored. The pile-up correction accounts for the occurrence of multiple collisions in a time span smaller than the detector resolution. The latter is of the order of 2\% in \pPb collisions and is negligible in \PbPb due to a lower collision rate. The final value is the average over the two methods. In \PbPb collisions, the normalization factor is computed for all the centrality classes considered.

In the p--Pb analysis, the invariant yield is multiplied by the MB cross section to obtain the Z-production cross section~\cite{ALICE-PUBLIC-2018-002}. In the Pb--Pb analysis, results are given both integrated and differential with respect to centrality and rapidity. The production is expressed as the invariant yield, normalized by the nuclear overlap function $\left< T_{\rm AA} \right>$. The centrality is expressed as $\left< \Npart \right>$, the average number of participant nucleons. The $\left< T_{\rm AA} \right>$ and $\left< \Npart \right>$ quantities are estimated via a Glauber model fit of the signal amplitude in the two arrays of the V0 detector~\cite{ALICE-PUBLIC-2018-011}. The nuclear modification of the production of a hard process, such as those producing the Z boson, is measured by \Raa, the ratio of the observed normalized yield in Pb--Pb collisions to that in pp collisions. Due to the insufficient integrated luminosity for pp collisions at \snn{} = 5 TeV, the pp reference is determined from pQCD theoretical calculations using the MCFM code with the CT14 PDF set~\cite{ct14}.

The relative systematic uncertainties for the p--Pb analysis are summarized in Table~\ref{table:systpPb}. The variation between the two methods for the computation of the normalization factor, which is less than 1\%, is taken as its systematic uncertainty. The evaluation of $A \times \epsilon$ is shown not to be affected by a change of PDF and nPDF set, or transport code in the MC simulations. The uncertainty on the Z-boson yield due to the tracking efficiency, evaluated to be 1\% (2\%) for the p-going (Pb-going) sample, is obtained by comparing the efficiency between data and MC, using the redundancy of the chambers of the tracking stations~\cite{Abelev:2014ffa}. The systematic uncertainty due to the dimuon trigger efficiency is determined by propagating the uncertainty on the efficiency of the detector elements, estimated with a data-driven method based on the redundancy of the trigger chamber information. The matching condition between tracks reconstructed in the tracking and triggering systems introduces a 1\% additional uncertainty. Finally, the systematic uncertainty associated to the alignment procedure is evaluated as the difference between the $A \times \epsilon$ computed with the data-driven tuning of the cluster resolution, with a global shift, and a MC parametrization without shift. This uncertainty is 7.7\% for the p-going dataset and 5.7\% for the Pb-going dataset, the difference between the two originating from the difference in the signal shape, which depends on rapidity. The total systematic uncertainty is determined by summing in quadrature the uncertainty from each source.

\begin{table}[t]
  \centering
  \caption{Components of the relative systematic uncertainties on the Z-boson yield and cross section in the p--Pb analysis. The uncertainties on the MB cross section and the alignment are partially correlated between p--Pb and Pb--p.}
\begin{tabular}{c|c}
  \hline
  Source                   & Relative systematic uncertainty (\%) \\
  \hline \hline
  Background contamination & 1.0 (p-going) \hspace{0.5cm} $<$ 0.1 (Pb-going) \\
  \hline
  Normalization factor     & 0.7 (p-going) \hspace{0.5cm} 0.2 (Pb-going) \\
  MB cross section         & 1.9 \\
  \hline
  Tracking efficiency      & 1.0 (p-going) \hspace{0.5cm} 2.0 (Pb-going) \\
  Trigger efficiency       & 1.0 \\
  Trigger/tracker matching & 1.0 \\
  Alignment                & 7.7 (p-going) \hspace{0.5cm} 5.7 (Pb-going) \\
  \hline
  Total                    & 8.2 (p-going) \hspace{0.5cm} 6.5 (Pb-going) \\
  \hline
\end{tabular}
\label{table:systpPb}
\end{table}

The sources and values of systematic uncertainties for the Pb--Pb analysis are displayed in Tab.~\ref{table:systPbPb}. The systematic uncertainties of the normalization factor, the tracking and trigger efficiencies, the trigger/tracker matching, and the alignment are evaluated in the same way as for the p--Pb analysis. The uncertainty of the centrality estimation and the average nuclear overlap function $\left< T_{\rm AA} \right>$ are obtained by varying the centrality class limits by $\pm$ 0.5\%, as detailed in Ref.~\cite{Abelev:2013qoq}. The uncertainty of the theoretical pp cross section, which is used as a reference for the \Raa computation, is obtained by varying the factorization and renormalization scales and accounting for the PDF uncertainty. This uncertainty is rapidity dependent and has values between 3.5\% and 5.0\%. The total systematic uncertainty is taken as the quadratic sum of all the sources.

\begin{table}[h]
  \centering
  \caption{Components of the relative systematic uncertainties on the Z-boson yield and $R_{\rm AA}$ in the Pb--Pb analysis. See text for details. The $\star$ symbol indicates a rapidity-dependent correlated uncertainty, while the uncertainty sources correlated as a function of centrality are marked by a $\diamond$. In the total lines are reported the cumulative systematic uncertainty of the result integrated in centrality and rapidity.}
\begin{tabular}{c|cc}
  \hline
  Source                    & Relative systematic uncertainty & (\%) \\
  \hline \hline
  Background contamination  & 1.0 & \\
  \hline
  Normalization factor      & 0.5 & $\star \diamond$ \\
  $\left< T_{\rm AA} \right>$ & 0.7 -- 1.5 & $\star$ \\
  Centrality estimation     & 0.2 -- 0.9 & $\star$ \\
  pp cross section        & 3.5 -- 5.0 & $\diamond$ \\
  \hline
  Tracking efficiency      & 3.0 & $\diamond$ \\
  Trigger efficiency       & 1.5 & $\diamond$ \\
  Trigger/tracker matching & 1.0 & $\diamond$ \\
  Alignment                & 5.0 & $\diamond$ \\
  \hline
  Total (yield)            & 6.3 \\
  Total ($R_{\rm AA}$)       & 7.4 \\
  \hline
\end{tabular}
\label{table:systPbPb}
\end{table}

\section{Results}
\label{sec:results}

\begin{figure}[h]
  \centering

    \includegraphics[width=0.6\textwidth]{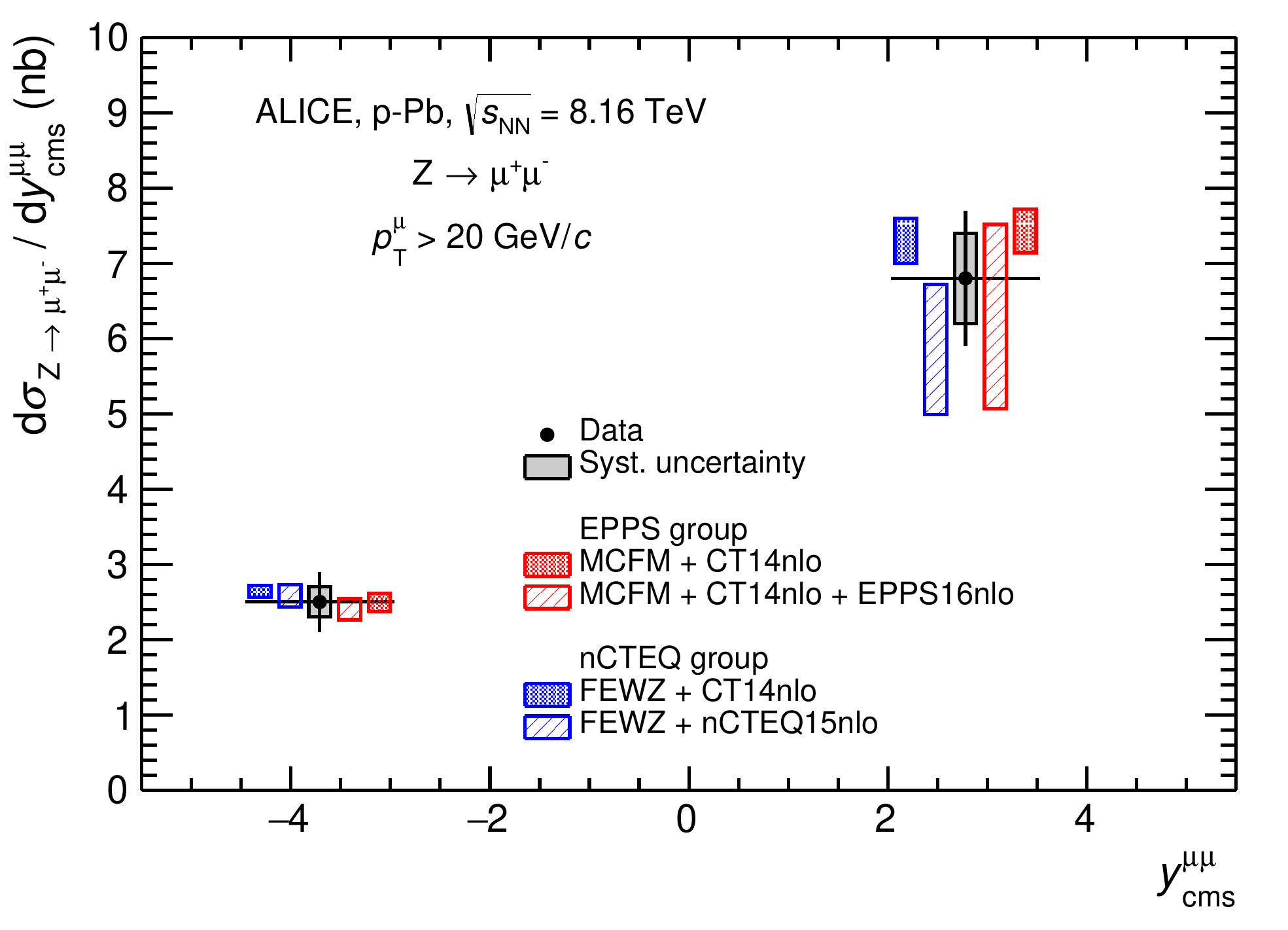}

    \caption{Production cross section of $\mu^+\mu^-$ from Z-boson decays, measured in p--Pb collisions at \eightnn and  compared  with theoretical calculations both based on CT14 (at NLO) free-nucleon PDFs~\cite{ct14} and on other PDF sets including the presence of a nuclear modification. The horizontal extension of the data points correspond to the measured rapidity range.  The bars and boxes correspond to the statistical and systematic uncertainties respectively. The theory points are horizontally shifted for better readability.}
  \label{fig:XsecVsTheory}
\end{figure}

The production cross section for the $\rm{Z} \rightarrow \mu^+\mu^-$ process  in p--Pb collisions at \eightnn with $\pt^{\mu} > 20$ \GeVc and $-4<\eta_\mu<-2.5$ is measured to be $\mathrm{d}\sigma^{\rm Pb-going}_{\rm{Z} \rightarrow \mu^+\mu^-}/\mathrm{d}y=2.5\pm 0.4\, (\mathrm{stat.})\,\pm 0.2\, (\mathrm{syst.})$ nb and $\mathrm{d}\sigma^{\rm p-going}_{\rm{Z} \rightarrow \mu^+\mu^-}/\mathrm{d}y=6.8\pm 0.9 \, (\mathrm{stat.})\, \pm 0.6\, (\mathrm{syst.})$  nb. 
In Fig.~\ref{fig:XsecVsTheory} the results
are compared with pQCD calculations with and without the nuclear modification of the parton distribution functions.
The Bjorken-$x$\linebreak range of partons in the Pb nucleus probed in the Pb-going collisions 
($-4.46<\ycms^{\mu\mu}<-2.96$)
is  above $10^{-1}$, while in p-going collisions
($2.03<\ycms^{\mu\mu}<3.53$)
it is roughly between  $10^{-4}$ and  $10^{-3}$. The former is expected to be mostly affected by antishadowing and EMC effects, while the latter is in the shadowing region.
The observed difference between the backward and forward cross sections is mainly due to the asymmetry of the collision
and is consistent with that predicted by  theoretical calculations for nucleon--nucleon collisions, as shown in the figure. The forward-$y$  region is closer to midrapidity where production cross sections are known to be larger.

The measurements are compared with two  model calculations based on pQCD at NLO. The first calculation utilizes the MCFM (Monte Carlo for FeMtobarn processes) code~\cite{Boughezal:2016wmq} using CT14 at NLO~\cite{ct14} as free-nucleon PDFs. The  EPPS16~\cite{Eskola:2016oht} parametrization of the nuclear modification to the PDFs is then considered to describe the lead environment. The second calculation uses the NNLO code FEWZ (Fully Exclusive W and Z Production)~\cite{Gavin:2010az}. The lead nucleus is modelled with nCTEQ15 nuclear PDFs~\cite{Kovarik:2015cma,Kusina:2016fxy}, while CT14 is used for the proton.  Both EPPS16 and nCTEQ15 rely on NLO calculations. The latter is a full nPDF set while EPPS16 is anchored to the CT14 free-nucleon PDFs.
More details on the  approximations and experimental datasets included in the extraction of the nPDFs can be found in Ref.~\cite{Paukkunen:2018kmm}. In all nuclear calculations, the proton and neutron contributions are weighted to reproduce the lead nucleus isospin.

Figure~\ref{fig:XsecVsTheory} shows that the measurements reported here are consistent with pQCD calculations incorporating both free-nucleon and nuclear-modified PDFs, within experimental and theoretical uncertainties. In p--Pb collisions the nuclear effects modify the parton distributions of only one of the two colliding nucleons and the inclusion of the nuclear modification of the PDFs results in a small change if compared to theoretical uncertainties. Moreover, the backward-$y$ region corresponds to a high Bjorken-$x$ range where multiple nuclear effects are present. These lead to both enhancement and depletion compared to free-nucleon PDFs. Their resulting effect is expected to be less pronounced than the one at forward-$y$ where only shadowing is present.

\begin{figure}[h]
  \centering
    \includegraphics[width=0.6\textwidth]{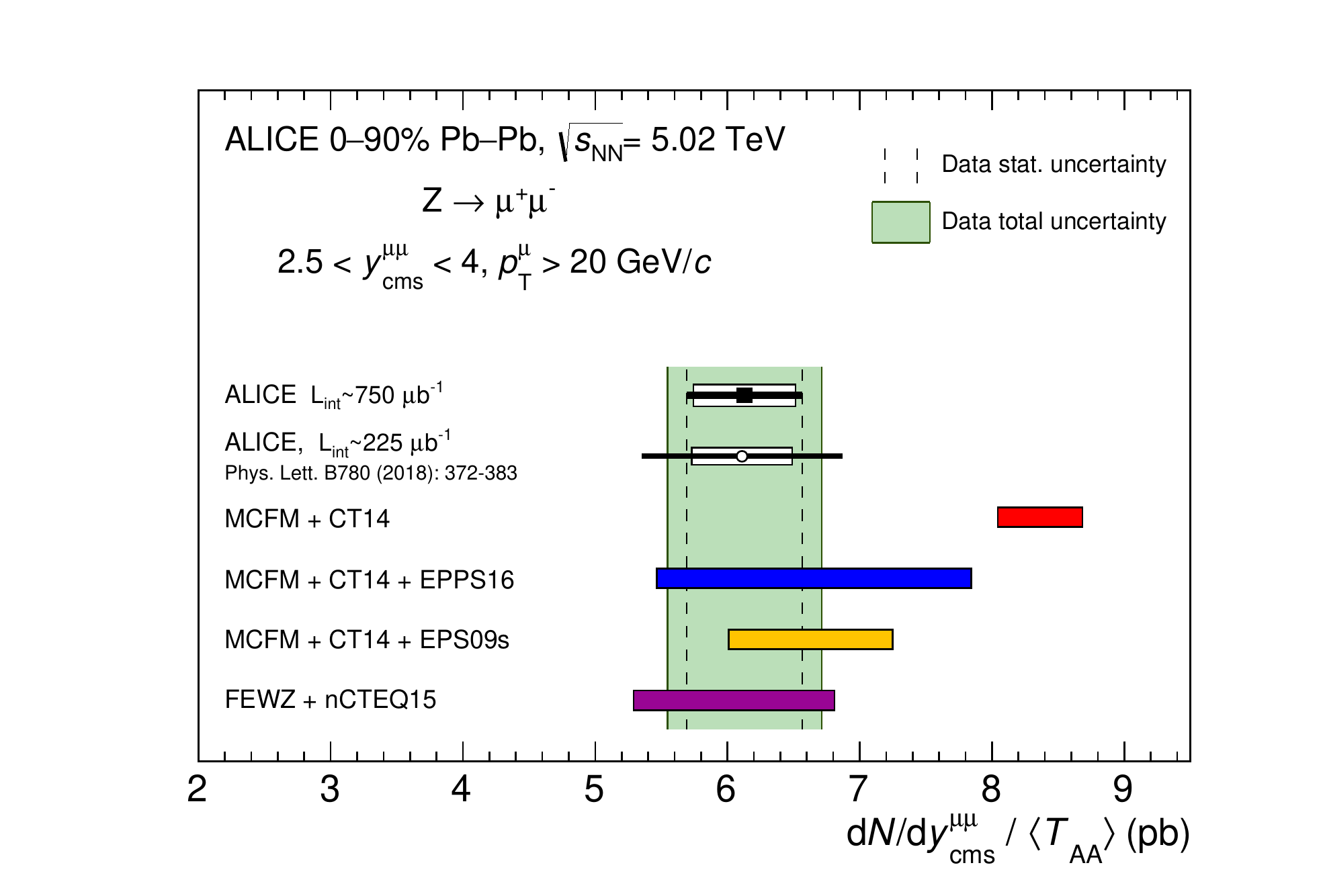}
  
  \caption{Invariant yield of $\mathrm{Z} \rightarrow \mu^+\mu^-$ divided by $\left< T_{\rm AA} \right>$ in the rapidity range $2.5 <\ycms^{\mu\mu} <4.0$ measured in Pb--Pb collisions at \fivenn in the centrality class 0--90\%. The vertical dashed band represents the statistical uncertainty of the data while the green filled band corresponds to the quadratic sum of statistical and systematic uncertainties. The result is compared with the previous ALICE result in the same collision system (the data of the earlier result are included in this analysis)~\cite{Acharya:2017wpf} , with CT14~\cite{ct14} free-nucleon PDF calculation and with several NLO pQCD calculations including nuclear modification of the PDFs. }
  \label{fig:IntegratedYield}
\end{figure}

The Z-boson invariant yield  normalized by the nuclear overlap function $\left< T_{\rm AA} \right>$ measured in Pb--Pb collisions at \fivenn  is
$6.1\pm 0.4\, (\mathrm{stat.})\,\pm 0.4\, (\mathrm{syst.})$ pb.
Because of the symmetry of the collision, the forward rapidity of this measurement probes simultaneously the high-$x$ and low-$x$ range of partons in the lead nucleus. As a result of the rapidity shift and the different nucleon--nucleon center-of-mass energy, these ranges are very close to those probed in \pPb.
In Fig.~\ref{fig:IntegratedYield} the normalized yield is compared with  the result previously published by ALICE~\cite{Acharya:2017wpf} based on the 2015 data sample which contains less than a third of the full statistics. The measurements are fully compatible with each other. The normalized yield is also compared with several pQCD calculations based on different codes (MCFM~\cite{Boughezal:2016wmq} or FEWZ~\cite{Gavin:2010az}) and different parton distribution sets. Along with CT14, CT14+EPPS16 and nCTEQ15 calculations~\cite{ct14, Eskola:2016oht,Kovarik:2015cma}, the calculation with CT14 baseline PDF and EPS09s nPDFs is also included in the comparison~\cite{Helenius:2012wd}. Although as a whole EPS09 is superseded by the more recent EPPS16, EPS09s is used because it contains a centrality dependence of the parton distributions which is not provided  in the EPPS16 nPDF set.
Neutron and proton contributions are properly weighted according to the lead isospin.
The uncertainties on the models include the uncertainty on the NLO calculations as well as the uncertainty on the parton distributions that are larger for those including nuclear effects. The large uncertainty of the EPPS16 calculation originates from the larger number of flavor degrees of freedom included in the parametrization~\cite{Paukkunen:2018kmm}.

The calculations using nuclear PDFs describe the yield measured in Pb--Pb collisions within uncertainties while the CT14-only calculation deviates from data by $3.4\sigma$ . This deviation is not observed in the p--Pb analysis for two main reasons. The first one is statistical. Although the Pb--Pb luminosity is smaller than the p--Pb one, the presence of more nuclear matter in Pb--Pb collisions makes the expected Z-boson yield greater than the one measured in the p--Pb samples, reducing the statistical uncertainty. Second, in Pb--Pb collisions, the distributions of both interacting partons experience nuclear modifications. In order to produce a Z boson at forward rapidity, a collision must occur between a low-$x$ and high-$x$ parton. This leads to a convolution of the shadowing effects at low $x$ and the net nuclear effect observed in backward-$y$ p--Pb collisions. Their combination enhances the suppression of the production with respect to what is separately measured in the two p--Pb rapidity regions.

At the moment most of the nPDF sets do not contain an explicit dependence on the position inside the nucleus, but they provide the average effect over all the nucleons in a given nucleus. Results which are fully inclusive in centrality can better accomodate in the global fitting procedure used to constrain the nPDFs. 
An estimation of  the integrated normalized invariant yield in the 0--100\% centrality interval is therefore important. Assuming that the yield scales with the number of nucleon--nucleon binary collisions \Ncoll, and with a conservative estimation of the nuclear modification in the 90--100\% centrality interval, the difference between the integrated normalized yields in 0--90\% and 0--100\% is found to be less than 1 per mille. This means that
the  present measurement can  be regarded also as the normalized invariant yield in the 0--100\% centrality interval given the current uncertainties.

\begin{figure}[h]
  \centering
  \subfigure{
    \includegraphics[width=0.48\textwidth]{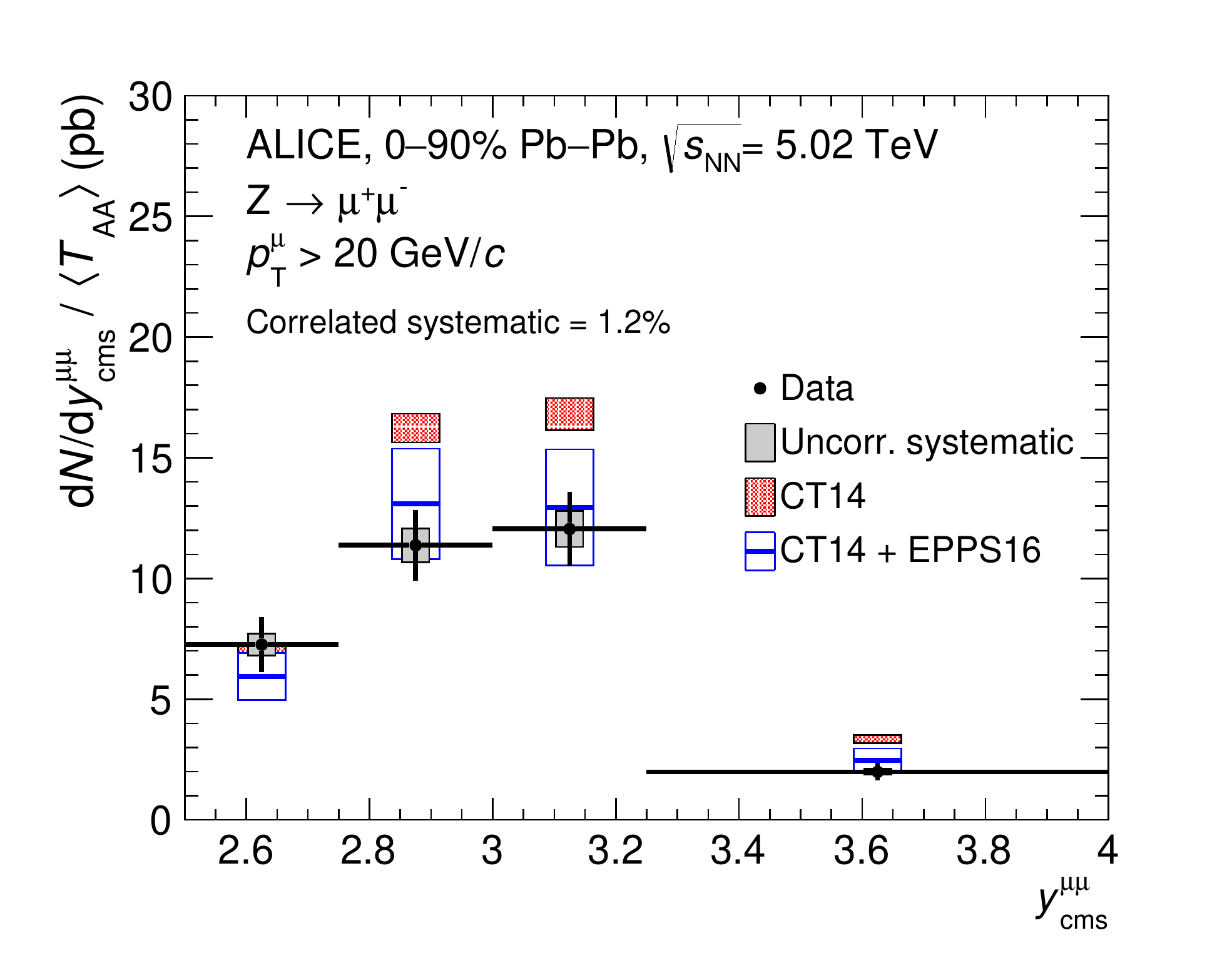}
  }
  \subfigure{
    \includegraphics[width=0.48\textwidth]{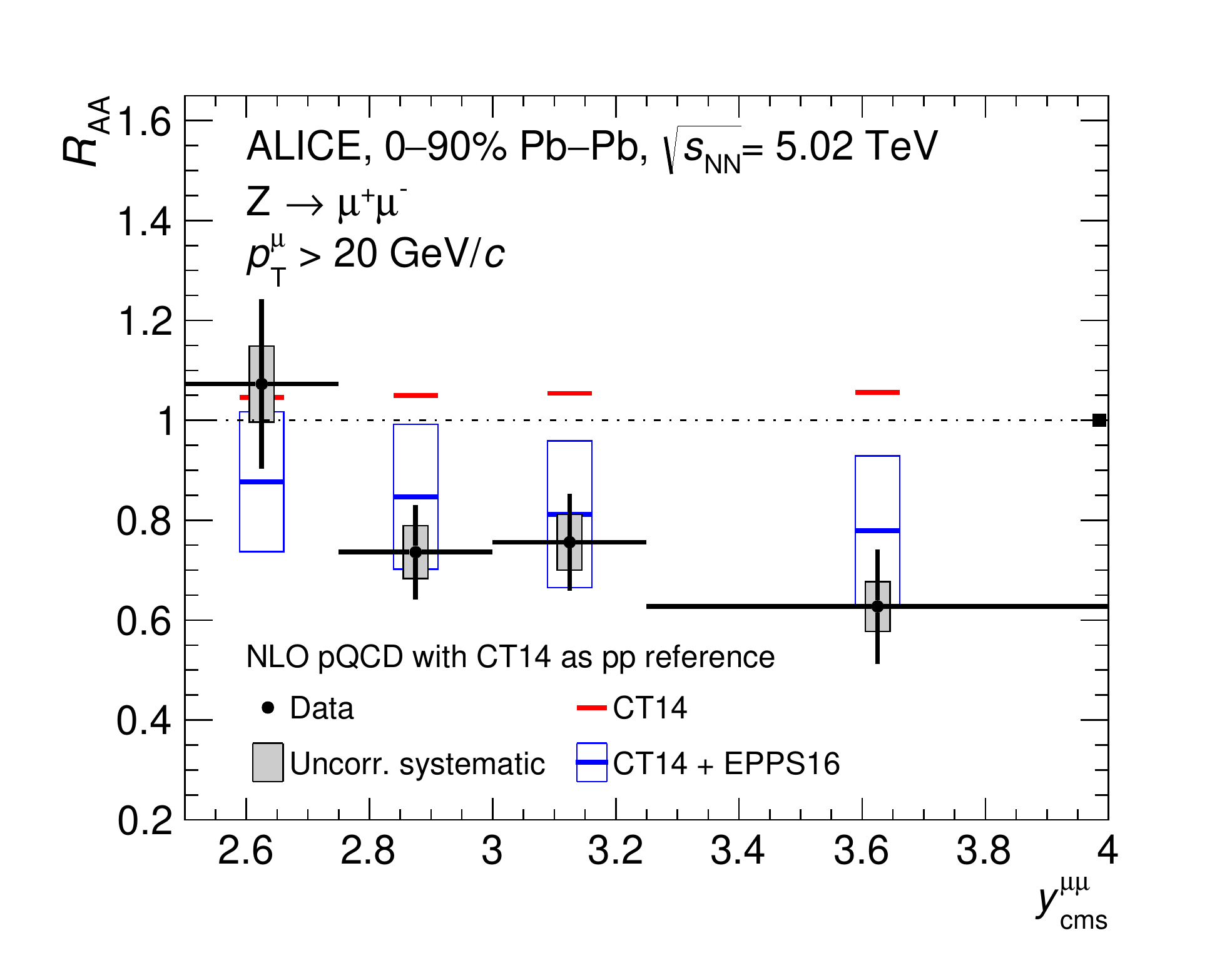}
  }
  \caption{Normalized invariant yield of $\mathrm{Z} \rightarrow \mu^+\mu^-$ (left panel) and corresponding \Raa (right panel) as a function of rapidity, measured in 0--90\%  Pb--Pb collisions at \fivenn. The vertical bars represent statistical uncertainties only. The horizontal extension corresponds to the rapidity bin width. The uncorrelated systematic uncertainties are reported as filled boxes. The \Raa correlated systematic uncertainty is displayed as a box on the unity line in the right panel. The CT14~\cite{ct14} (at NLO) pQCD proton--proton cross section is used as reference to compute \Raa. The results are compared with free-nucleon PDF (CT14) and with nuclear PDF (CT14+EPPS16~\cite{Eskola:2016oht}) calculations. The free-nucleon PDF calculations are larger than unity as a consequence of the isospin effect, which is properly taken into account by all the calculations.  }
  \label{fig:5TeVvsRapidity}
\end{figure}

The Z-boson production in Pb--Pb is also  studied as a function of rapidity and centrality. The left panel of Fig.~\ref{fig:5TeVvsRapidity} shows the normalized invariant yield in the rapidity intervals $2.5<\ycms^{\mu\mu}<2.75$, $2.75<\ycms^{\mu\mu}<3$, $3<\ycms^{\mu\mu}<3.25$ and $3.25<\ycms^{\mu\mu}<4$. The results are compared with CT14 predictions both with and without EPPS16 nuclear modification. A shadowing effect is foreseen in the full rapidity range. The right panel of Fig.~\ref{fig:5TeVvsRapidity} shows the rapidity dependence of the nuclear modification factor \Raa, computed by dividing the yield normalized to $\left< T_{\rm AA} \right>$  by the pp cross section at \five obtained with pQCD calculations with CT14 PDFs. For this observable, the uncertainties on the free-nucleon PDFs are factored out in the theoretical calculations, and the remaining uncorrelated uncertainties are related to the nuclear PDFs only. The measured \Raa is in agreement within uncertainties with the EPPS16 calculations while, at large rapidity, it deviates from the free-nucleon calculations.

\begin{figure}[h]
  \centering
    \includegraphics[width=0.48\linewidth]{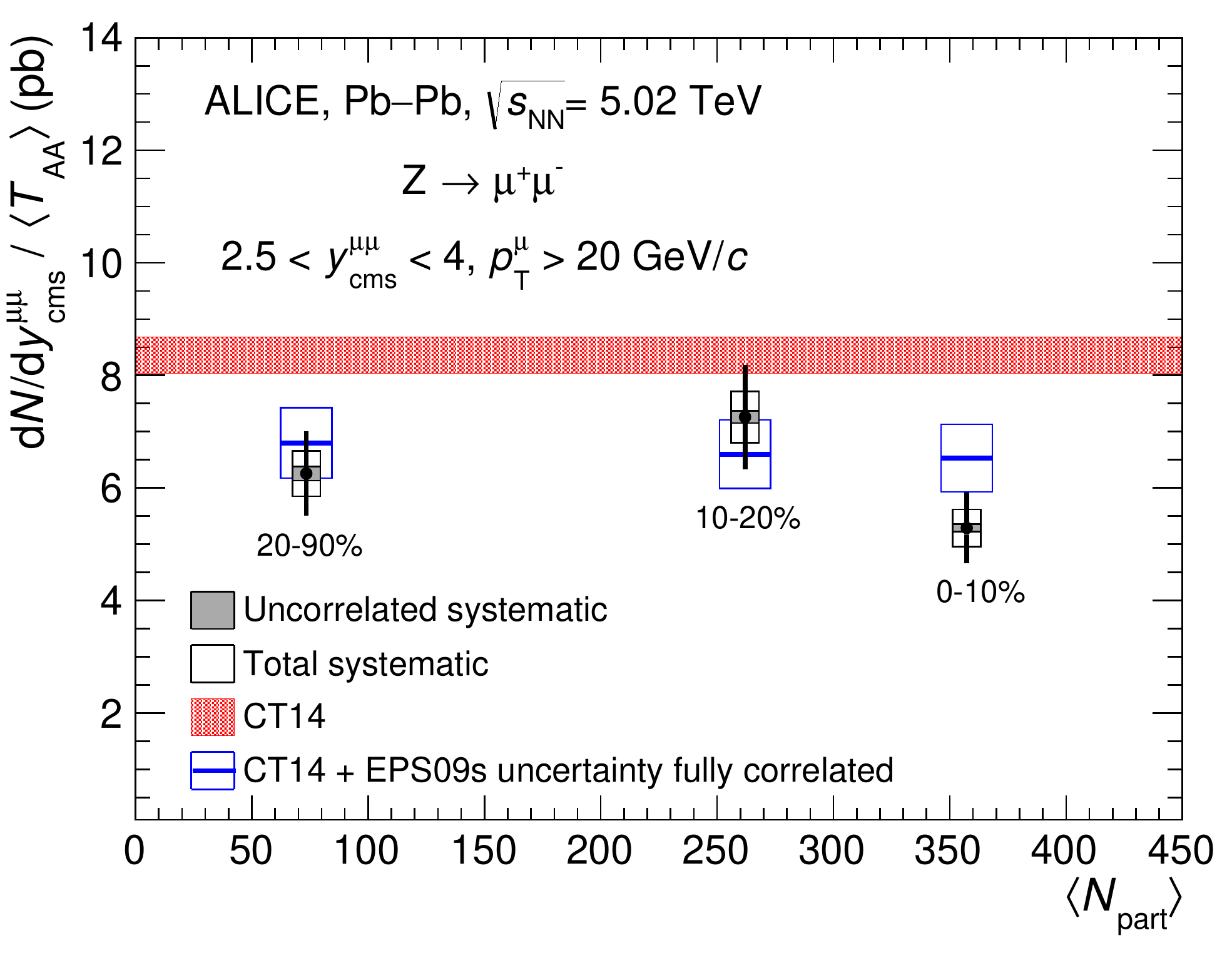}
  
  \caption{Invariant yield of $\mathrm{Z} \rightarrow \mu^+\mu^-$ divided by $\left< T_{\rm AA} \right>$ in the rapidity range $2.5 <\ycms^{\mu\mu} <4.0$ for three centrality classes in Pb--Pb collisions at \fivenn. The uncorrelated systematic uncertainties are reported as filled boxes. The open boxes indicate the quadratic sum of correlated and uncorrelated systematics. The results are compared with the free-nucleon PDF prediction (CT14~\cite{ct14}) and with calculations with the centrality-dependent EPS09s nPDFs ~\cite{Helenius:2012wd}. }
  \label{fig:5TeVvsCentrality}
\end{figure}

In Fig.~\ref{fig:5TeVvsCentrality}, the normalized invariant yield is shown as a function of centrality. The CT14 calculations are based on free-nucleon PDFs and therefore, by construction, carry no centrality dependence. The data are also compared with calculations from EPS09s~\cite{Helenius:2012wd}, which show a decrease in the invariant yield towards more central collisions, although the effect is very weak. 
Furthermore, in each centrality bin the EPS09s prediction is consistent with the more recent EPPS16 set, which does not implement a dependence on the impact parameter (the CT14+EPPS16 calculation is displayed in Fig.~\ref{fig:IntegratedYield}).

Within uncertainties, each data point is well described by models including nPDFs, while the CT14-only calculation overestimates the data, especially for the most central collisions where the difference is $3.9\sigma$.

\section{Conclusions}
\label{sec:conclusion}

The Z-boson production has been studied at large rapidities in \pPb collisions at \eightnn and in \PbPb collisions at \fivenn. 

For the \pPb collisions, the Z bosons were measured in the rapidity range $-4.46 < \ycms^{\mu\mu} < -2.96$ and $2.03 < \ycms^{\mu\mu} < 3.53$. The production cross section at forward and backward rapidity has been compared with theoretical predictions, both with and without nuclear modifications. The data show little sensitivity to the presence of nuclear effects, partially because in \pPb collisions, nuclear modifications to the PDFs affect only one of the two colliding particles. This is particularly true in the backward region, where enhancement and depletion effects on nPDFs tend to compensate. As a result, the calculations for the nuclear modification of the PDFs are very close to those without. In the forward region, low-$x$ partons of the Pb nucleus are probed which are only sensitive to shadowing (which corresponds to a depletion in the nPDFs). Consequently, nuclear effects tend more clearly to induce a decrease in the cross section. Nonetheless it remains compatible within uncertainties with the one calculated neglecting such effects.

In the Pb--Pb data, the invariant yield normalized by the average nuclear overlap function has
been evaluated in the rapidity range from $2.5 < \ycms^{\mu\mu} < 4$ and in the 0--90\% centrality class. The results obtained in this paper supersede those from an earlier ALICE publication~\cite{Acharya:2017wpf}, where only part of the current dataset was used. The experimental data are, within uncertainties, in agreement with theoretical calculations that include various parametrizations of nuclear modification of the PDFs.
On the contrary, the integrated yield deviates by $3.4\sigma$ from the prediction obtained using free-nucleon PDFs.

Comparisons with models of the measured differential yields versus centrality and rapidity were also carried out, generally showing agreement with nuclear modified PDFs. In contrast, a discrepancy with calculations based on free-nucleon PDFs was found. The differential measurements presented in this paper can provide additional constraints to the nPDFs.

%%%%%%%%%%%%%%%%%%%%%%%%%%%%%%%%
% end main text 
%%%%%%%%%%%%%%%%%%%%%%%%%%%%%%%%

%%%%% acknowledgements - handled by EB chairs 
\newenvironment{acknowledgement}{\relax}{\relax}
\begin{acknowledgement}
\section*{Acknowledgements}
% add specific acknowledgements here 
% ...but please don't remove the line below: funding agencies
% will be acknowledged with a custom tex file handled by EB chairs after Collab Round 2
The authors would like to extend special thanks to H. Paukkunen and I. Helenius for providing the CT14NLO, EPS09s and EPPS16 calculations, as well as K. Kovarik, A. Kusina, F. Olness, I. Schienbein and T. Tunks for the nCTEQ predictions.
% Version: 2020-05-08

The ALICE Collaboration would like to thank all its engineers and technicians for their invaluable contributions to the construction of the experiment and the CERN accelerator teams for the outstanding performance of the LHC complex.
The ALICE Collaboration gratefully acknowledges the resources and support provided by all Grid centres and the Worldwide LHC Computing Grid (WLCG) collaboration.
The ALICE Collaboration acknowledges the following funding agencies for their support in building and running the ALICE detector:
A. I. Alikhanyan National Science Laboratory (Yerevan Physics Institute) Foundation (ANSL), State Committee of Science and World Federation of Scientists (WFS), Armenia;
Austrian Academy of Sciences, Austrian Science Fund (FWF): [M 2467-N36] and Nationalstiftung f\"{u}r Forschung, Technologie und Entwicklung, Austria;
Ministry of Communications and High Technologies, National Nuclear Research Center, Azerbaijan;
Conselho Nacional de Desenvolvimento Cient\'{\i}fico e Tecnol\'{o}gico (CNPq), Financiadora de Estudos e Projetos (Finep), Funda\c{c}\~{a}o de Amparo \`{a} Pesquisa do Estado de S\~{a}o Paulo (FAPESP) and Universidade Federal do Rio Grande do Sul (UFRGS), Brazil;
Ministry of Education of China (MOEC) , Ministry of Science \& Technology of China (MSTC) and National Natural Science Foundation of China (NSFC), China;
Ministry of Science and Education and Croatian Science Foundation, Croatia;
Centro de Aplicaciones Tecnol\'{o}gicas y Desarrollo Nuclear (CEADEN), Cubaenerg\'{\i}a, Cuba;
Ministry of Education, Youth and Sports of the Czech Republic, Czech Republic;
The Danish Council for Independent Research | Natural Sciences, the VILLUM FONDEN and Danish National Research Foundation (DNRF), Denmark;
Helsinki Institute of Physics (HIP), Finland;
Commissariat \`{a} l'Energie Atomique (CEA) and Institut National de Physique Nucl\'{e}aire et de Physique des Particules (IN2P3) and Centre National de la Recherche Scientifique (CNRS), France;
Bundesministerium f\"{u}r Bildung und Forschung (BMBF) and GSI Helmholtzzentrum f\"{u}r Schwerionenforschung GmbH, Germany;
General Secretariat for Research and Technology, Ministry of Education, Research and Religions, Greece;
National Research, Development and Innovation Office, Hungary;
Department of Atomic Energy Government of India (DAE), Department of Science and Technology, Government of India (DST), University Grants Commission, Government of India (UGC) and Council of Scientific and Industrial Research (CSIR), India;
Indonesian Institute of Science, Indonesia;
Centro Fermi - Museo Storico della Fisica e Centro Studi e Ricerche Enrico Fermi and Istituto Nazionale di Fisica Nucleare (INFN), Italy;
Institute for Innovative Science and Technology , Nagasaki Institute of Applied Science (IIST), Japanese Ministry of Education, Culture, Sports, Science and Technology (MEXT) and Japan Society for the Promotion of Science (JSPS) KAKENHI, Japan;
Consejo Nacional de Ciencia (CONACYT) y Tecnolog\'{i}a, through Fondo de Cooperaci\'{o}n Internacional en Ciencia y Tecnolog\'{i}a (FONCICYT) and Direcci\'{o}n General de Asuntos del Personal Academico (DGAPA), Mexico;
Nederlandse Organisatie voor Wetenschappelijk Onderzoek (NWO), Netherlands;
The Research Council of Norway, Norway;
Commission on Science and Technology for Sustainable Development in the South (COMSATS), Pakistan;
Pontificia Universidad Cat\'{o}lica del Per\'{u}, Peru;
Ministry of Science and Higher Education, National Science Centre and WUT ID-UB, Poland;
Korea Institute of Science and Technology Information and National Research Foundation of Korea (NRF), Republic of Korea;
Ministry of Education and Scientific Research, Institute of Atomic Physics and Ministry of Research and Innovation and Institute of Atomic Physics, Romania;
Joint Institute for Nuclear Research (JINR), Ministry of Education and Science of the Russian Federation, National Research Centre Kurchatov Institute, Russian Science Foundation and Russian Foundation for Basic Research, Russia;
Ministry of Education, Science, Research and Sport of the Slovak Republic, Slovakia;
National Research Foundation of South Africa, South Africa;
Swedish Research Council (VR) and Knut \& Alice Wallenberg Foundation (KAW), Sweden;
European Organization for Nuclear Research, Switzerland;
Suranaree University of Technology (SUT), National Science and Technology Development Agency (NSDTA) and Office of the Higher Education Commission under NRU project of Thailand, Thailand;
Turkish Atomic Energy Agency (TAEK), Turkey;
National Academy of  Sciences of Ukraine, Ukraine;
Science and Technology Facilities Council (STFC), United Kingdom;
National Science Foundation of the United States of America (NSF) and United States Department of Energy, Office of Nuclear Physics (DOE NP), United States of America.
\end{acknowledgement}

% %%%%%%%% Bibliography 
\bibliographystyle{utphys}   % Remember we use title in the biblio
\bibliography{bibliography}

%%%%%%%%%%%%%%%%%%%%%%%%%%%%%%%%
% Appendices: yours (if any) + authorlist
%%%%%%%%%%%%%%%%%%%%%%%%%%%%%%%%
\newpage
\appendix

%
%\input{appendix.tex} % put your appendices here (if any)
%

%%%%% Authorlist - please do not touch: handled by EB chairs 
\section{The ALICE Collaboration}
\label{app:collab}
% Collaboration: CERN-LHC-ALICE
% Generation Date is 2020-05-08

% How to use:
%%%%%%%%% appendix with author list
%\appendix
%\section{The ALICE Collaboration}
%\label{app:collab}
%\input{Alice_Authorslist_XXXX-Axx-XX.tex}
\begingroup
\small
\begin{flushleft}
S.~Acharya\Irefn{org141}\And 
D.~Adamov\'{a}\Irefn{org95}\And 
A.~Adler\Irefn{org74}\And 
J.~Adolfsson\Irefn{org81}\And 
M.M.~Aggarwal\Irefn{org100}\And 
G.~Aglieri Rinella\Irefn{org34}\And 
M.~Agnello\Irefn{org30}\And 
N.~Agrawal\Irefn{org10}\textsuperscript{,}\Irefn{org54}\And 
Z.~Ahammed\Irefn{org141}\And 
S.~Ahmad\Irefn{org16}
% ...
\And
\renewcommand\labelenumi{\textsuperscript{\theenumi}~}

\section*{Affiliation notes}
\renewcommand\theenumi{\roman{enumi}}
\begin{Authlist}
\item \Adef{org*}Deceased
\item \Adef{orgI}Italian National Agency for New Technologies, Energy and Sustainable Economic Development (ENEA), Bologna, Italy
\item \Adef{orgII}Dipartimento DET del Politecnico di Torino, Turin, Italy
\item \Adef{orgIII}M.V. Lomonosov Moscow State University, D.V. Skobeltsyn Institute of Nuclear, Physics, Moscow, Russia
\item \Adef{orgIV}Department of Applied Physics, Aligarh Muslim University, Aligarh, India
\item \Adef{orgV}Institute of Theoretical Physics, University of Wroclaw, Poland
\end{Authlist}

\section*{Collaboration Institutes}
\renewcommand\theenumi{\arabic{enumi}~}
\begin{Authlist}
\item \Idef{org1}A.I. Alikhanyan National Science Laboratory (Yerevan Physics Institute) Foundation, Yerevan, Armenia
\item \Idef{org2}Bogolyubov Institute for Theoretical Physics, National Academy of Sciences of Ukraine, Kiev, Ukraine
\item \Idef{org3}Bose Institute, Department of Physics  and Centre for Astroparticle Physics and Space Science (CAPSS), Kolkata, India
\item \Idef{org4}Budker Institute for Nuclear Physics, Novosibirsk, Russia
\item \Idef{org5}California Polytechnic State University, San Luis Obispo, California, United States
\item \Idef{org6}Central China Normal University, Wuhan, China
\item \Idef{org7}Centre de Calcul de l'IN2P3, Villeurbanne, Lyon, France
\item \Idef{org8}Centro de Aplicaciones Tecnol\'{o}gicas y Desarrollo Nuclear (CEADEN), Havana, Cuba
\item \Idef{org9}Centro de Investigaci\'{o}n y de Estudios Avanzados (CINVESTAV), Mexico City and M\'{e}rida, Mexico
\item \Idef{org10}Centro Fermi - Museo Storico della Fisica e Centro Studi e Ricerche ``Enrico Fermi', Rome, Italy
\item \Idef{org11}Chicago State University, Chicago, Illinois, United States
\item \Idef{org12}China Institute of Atomic Energy, Beijing, China
\item \Idef{org13}Comenius University Bratislava, Faculty of Mathematics, Physics and Informatics, Bratislava, Slovakia
\item \Idef{org14}COMSATS University Islamabad, Islamabad, Pakistan
\item \Idef{org15}Creighton University, Omaha, Nebraska, United States
\item \Idef{org16}Department of Physics, Aligarh Muslim University, Aligarh, India
\item \Idef{org17}Department of Physics, Pusan National University, Pusan, Republic of Korea
\item \Idef{org18}Department of Physics, Sejong University, Seoul, Republic of Korea
\item \Idef{org19}Department of Physics, University of California, Berkeley, California, United States
\item \Idef{org20}Department of Physics, University of Oslo, Oslo, Norway
\item \Idef{org21}Department of Physics and Technology, University of Bergen, Bergen, Norway
\item \Idef{org22}Dipartimento di Fisica dell'Universit\`{a} 'La Sapienza' and Sezione INFN, Rome, Italy
\item \Idef{org23}Dipartimento di Fisica dell'Universit\`{a} and Sezione INFN, Cagliari, Italy
\item \Idef{org24}Dipartimento di Fisica dell'Universit\`{a} and Sezione INFN, Trieste, Italy
\item \Idef{org25}Dipartimento di Fisica dell'Universit\`{a} and Sezione INFN, Turin, Italy
\item \Idef{org26}Dipartimento di Fisica e Astronomia dell'Universit\`{a} and Sezione INFN, Bologna, Italy
\item \Idef{org27}Dipartimento di Fisica e Astronomia dell'Universit\`{a} and Sezione INFN, Catania, Italy
\item \Idef{org28}Dipartimento di Fisica e Astronomia dell'Universit\`{a} and Sezione INFN, Padova, Italy
\item \Idef{org29}Dipartimento di Fisica `E.R.~Caianiello' dell'Universit\`{a} and Gruppo Collegato INFN, Salerno, Italy
\item \Idef{org30}Dipartimento DISAT del Politecnico and Sezione INFN, Turin, Italy
\item \Idef{org31}Dipartimento di Scienze e Innovazione Tecnologica dell'Universit\`{a} del Piemonte Orientale and INFN Sezione di Torino, Alessandria, Italy
\item \Idef{org32}Dipartimento di Scienze MIFT, Universit\`{a} di Messina, Messina, Italy
\item \Idef{org33}Dipartimento Interateneo di Fisica `M.~Merlin' and Sezione INFN, Bari, Italy
\item \Idef{org34}European Organization for Nuclear Research (CERN), Geneva, Switzerland
\item \Idef{org35}Faculty of Electrical Engineering, Mechanical Engineering and Naval Architecture, University of Split, Split, Croatia
\item \Idef{org36}Faculty of Engineering and Science, Western Norway University of Applied Sciences, Bergen, Norway
\item \Idef{org37}Faculty of Nuclear Sciences and Physical Engineering, Czech Technical University in Prague, Prague, Czech Republic
\item \Idef{org38}Faculty of Science, P.J.~\v{S}af\'{a}rik University, Ko\v{s}ice, Slovakia
\item \Idef{org39}Frankfurt Institute for Advanced Studies, Johann Wolfgang Goethe-Universit\"{a}t Frankfurt, Frankfurt, Germany
\item \Idef{org40}Fudan University, Shanghai, China
\item \Idef{org41}Gangneung-Wonju National University, Gangneung, Republic of Korea
\item \Idef{org42}Gauhati University, Department of Physics, Guwahati, India
\item \Idef{org43}Helmholtz-Institut f\"{u}r Strahlen- und Kernphysik, Rheinische Friedrich-Wilhelms-Universit\"{a}t Bonn, Bonn, Germany
\item \Idef{org44}Helsinki Institute of Physics (HIP), Helsinki, Finland
\item \Idef{org45}High Energy Physics Group,  Universidad Aut\'{o}noma de Puebla, Puebla, Mexico
\item \Idef{org46}Hiroshima University, Hiroshima, Japan
\item \Idef{org47}Hochschule Worms, Zentrum  f\"{u}r Technologietransfer und Telekommunikation (ZTT), Worms, Germany
\item \Idef{org48}Horia Hulubei National Institute of Physics and Nuclear Engineering, Bucharest, Romania
\item \Idef{org49}Indian Institute of Technology Bombay (IIT), Mumbai, India
\item \Idef{org50}Indian Institute of Technology Indore, Indore, India
\item \Idef{org51}Indonesian Institute of Sciences, Jakarta, Indonesia
\item \Idef{org52}INFN, Laboratori Nazionali di Frascati, Frascati, Italy
\item \Idef{org53}INFN, Sezione di Bari, Bari, Italy
\item \Idef{org54}INFN, Sezione di Bologna, Bologna, Italy
\item \Idef{org55}INFN, Sezione di Cagliari, Cagliari, Italy
\item \Idef{org56}INFN, Sezione di Catania, Catania, Italy
\item \Idef{org57}INFN, Sezione di Padova, Padova, Italy
\item \Idef{org58}INFN, Sezione di Roma, Rome, Italy
\item \Idef{org59}INFN, Sezione di Torino, Turin, Italy
\item \Idef{org60}INFN, Sezione di Trieste, Trieste, Italy
\item \Idef{org61}Inha University, Incheon, Republic of Korea
\item \Idef{org62}Institute for Nuclear Research, Academy of Sciences, Moscow, Russia
\item \Idef{org63}Institute for Subatomic Physics, Utrecht University/Nikhef, Utrecht, Netherlands
\item \Idef{org64}Institute of Experimental Physics, Slovak Academy of Sciences, Ko\v{s}ice, Slovakia
\item \Idef{org65}Institute of Physics, Homi Bhabha National Institute, Bhubaneswar, India
\item \Idef{org66}Institute of Physics of the Czech Academy of Sciences, Prague, Czech Republic
\item \Idef{org67}Institute of Space Science (ISS), Bucharest, Romania
\item \Idef{org68}Institut f\"{u}r Kernphysik, Johann Wolfgang Goethe-Universit\"{a}t Frankfurt, Frankfurt, Germany
\item \Idef{org69}Instituto de Ciencias Nucleares, Universidad Nacional Aut\'{o}noma de M\'{e}xico, Mexico City, Mexico
\item \Idef{org70}Instituto de F\'{i}sica, Universidade Federal do Rio Grande do Sul (UFRGS), Porto Alegre, Brazil
\item \Idef{org71}Instituto de F\'{\i}sica, Universidad Nacional Aut\'{o}noma de M\'{e}xico, Mexico City, Mexico
\item \Idef{org72}iThemba LABS, National Research Foundation, Somerset West, South Africa
\item \Idef{org73}Jeonbuk National University, Jeonju, Republic of Korea
\item \Idef{org74}Johann-Wolfgang-Goethe Universit\"{a}t Frankfurt Institut f\"{u}r Informatik, Fachbereich Informatik und Mathematik, Frankfurt, Germany
\item \Idef{org75}Joint Institute for Nuclear Research (JINR), Dubna, Russia
\item \Idef{org76}Korea Institute of Science and Technology Information, Daejeon, Republic of Korea
\item \Idef{org77}KTO Karatay University, Konya, Turkey
\item \Idef{org78}Laboratoire de Physique des 2 Infinis, Ir\`{e}ne Joliot-Curie, Orsay, France
\item \Idef{org79}Laboratoire de Physique Subatomique et de Cosmologie, Universit\'{e} Grenoble-Alpes, CNRS-IN2P3, Grenoble, France
\item \Idef{org80}Lawrence Berkeley National Laboratory, Berkeley, California, United States
\item \Idef{org81}Lund University Department of Physics, Division of Particle Physics, Lund, Sweden
\item \Idef{org82}Nagasaki Institute of Applied Science, Nagasaki, Japan
\item \Idef{org83}Nara Women{'}s University (NWU), Nara, Japan
\item \Idef{org84}National and Kapodistrian University of Athens, School of Science, Department of Physics , Athens, Greece
\item \Idef{org85}National Centre for Nuclear Research, Warsaw, Poland
\item \Idef{org86}National Institute of Science Education and Research, Homi Bhabha National Institute, Jatni, India
\item \Idef{org87}National Nuclear Research Center, Baku, Azerbaijan
\item \Idef{org88}National Research Centre Kurchatov Institute, Moscow, Russia
\item \Idef{org89}Niels Bohr Institute, University of Copenhagen, Copenhagen, Denmark
\item \Idef{org90}Nikhef, National institute for subatomic physics, Amsterdam, Netherlands
\item \Idef{org91}NRC Kurchatov Institute IHEP, Protvino, Russia
\item \Idef{org92}NRC \guillemotleft Kurchatov\guillemotright~Institute - ITEP, Moscow, Russia
\item \Idef{org93}NRNU Moscow Engineering Physics Institute, Moscow, Russia
\item \Idef{org94}Nuclear Physics Group, STFC Daresbury Laboratory, Daresbury, United Kingdom
\item \Idef{org95}Nuclear Physics Institute of the Czech Academy of Sciences, \v{R}e\v{z} u Prahy, Czech Republic
\item \Idef{org96}Oak Ridge National Laboratory, Oak Ridge, Tennessee, United States
\item \Idef{org97}Ohio State University, Columbus, Ohio, United States
\item \Idef{org98}Petersburg Nuclear Physics Institute, Gatchina, Russia
\item \Idef{org99}Physics department, Faculty of science, University of Zagreb, Zagreb, Croatia
\item \Idef{org100}Physics Department, Panjab University, Chandigarh, India
\item \Idef{org101}Physics Department, University of Jammu, Jammu, India
\item \Idef{org102}Physics Department, University of Rajasthan, Jaipur, India
\item \Idef{org103}Physikalisches Institut, Eberhard-Karls-Universit\"{a}t T\"{u}bingen, T\"{u}bingen, Germany
\item \Idef{org104}Physikalisches Institut, Ruprecht-Karls-Universit\"{a}t Heidelberg, Heidelberg, Germany
\item \Idef{org105}Physik Department, Technische Universit\"{a}t M\"{u}nchen, Munich, Germany
\item \Idef{org106}Politecnico di Bari, Bari, Italy
\item \Idef{org107}Research Division and ExtreMe Matter Institute EMMI, GSI Helmholtzzentrum f\"ur Schwerionenforschung GmbH, Darmstadt, Germany
\item \Idef{org108}Rudjer Bo\v{s}kovi\'{c} Institute, Zagreb, Croatia
\item \Idef{org109}Russian Federal Nuclear Center (VNIIEF), Sarov, Russia
\item \Idef{org110}Saha Institute of Nuclear Physics, Homi Bhabha National Institute, Kolkata, India
\item \Idef{org111}School of Physics and Astronomy, University of Birmingham, Birmingham, United Kingdom
\item \Idef{org112}Secci\'{o}n F\'{\i}sica, Departamento de Ciencias, Pontificia Universidad Cat\'{o}lica del Per\'{u}, Lima, Peru
\item \Idef{org113}St. Petersburg State University, St. Petersburg, Russia
\item \Idef{org114}Stefan Meyer Institut f\"{u}r Subatomare Physik (SMI), Vienna, Austria
\item \Idef{org115}SUBATECH, IMT Atlantique, Universit\'{e} de Nantes, CNRS-IN2P3, Nantes, France
\item \Idef{org116}Suranaree University of Technology, Nakhon Ratchasima, Thailand
\item \Idef{org117}Technical University of Ko\v{s}ice, Ko\v{s}ice, Slovakia
\item \Idef{org118}The Henryk Niewodniczanski Institute of Nuclear Physics, Polish Academy of Sciences, Cracow, Poland
\item \Idef{org119}The University of Texas at Austin, Austin, Texas, United States
\item \Idef{org120}Universidad Aut\'{o}noma de Sinaloa, Culiac\'{a}n, Mexico
\item \Idef{org121}Universidade de S\~{a}o Paulo (USP), S\~{a}o Paulo, Brazil
\item \Idef{org122}Universidade Estadual de Campinas (UNICAMP), Campinas, Brazil
\item \Idef{org123}Universidade Federal do ABC, Santo Andre, Brazil
\item \Idef{org124}University of Cape Town, Cape Town, South Africa
\item \Idef{org125}University of Houston, Houston, Texas, United States
\item \Idef{org126}University of Jyv\"{a}skyl\"{a}, Jyv\"{a}skyl\"{a}, Finland
\item \Idef{org127}University of Liverpool, Liverpool, United Kingdom
\item \Idef{org128}University of Science and Technology of China, Hefei, China
\item \Idef{org129}University of South-Eastern Norway, Tonsberg, Norway
\item \Idef{org130}University of Tennessee, Knoxville, Tennessee, United States
\item \Idef{org131}University of the Witwatersrand, Johannesburg, South Africa
\item \Idef{org132}University of Tokyo, Tokyo, Japan
\item \Idef{org133}University of Tsukuba, Tsukuba, Japan
\item \Idef{org134}Universit\'{e} Clermont Auvergne, CNRS/IN2P3, LPC, Clermont-Ferrand, France
\item \Idef{org135}Universit\'{e} de Lyon, Universit\'{e} Lyon 1, CNRS/IN2P3, IPN-Lyon, Villeurbanne, Lyon, France
\item \Idef{org136}Universit\'{e} de Strasbourg, CNRS, IPHC UMR 7178, F-67000 Strasbourg, France, Strasbourg, France
\item \Idef{org137}Universit\'{e} Paris-Saclay Centre d'Etudes de Saclay (CEA), IRFU, D\'{e}partment de Physique Nucl\'{e}aire (DPhN), Saclay, France
\item \Idef{org138}Universit\`{a} degli Studi di Foggia, Foggia, Italy
\item \Idef{org139}Universit\`{a} degli Studi di Pavia, Pavia, Italy
\item \Idef{org140}Universit\`{a} di Brescia, Brescia, Italy
\item \Idef{org141}Variable Energy Cyclotron Centre, Homi Bhabha National Institute, Kolkata, India
\item \Idef{org142}Warsaw University of Technology, Warsaw, Poland
\item \Idef{org143}Wayne State University, Detroit, Michigan, United States
\item \Idef{org144}Westf\"{a}lische Wilhelms-Universit\"{a}t M\"{u}nster, Institut f\"{u}r Kernphysik, M\"{u}nster, Germany
\item \Idef{org145}Wigner Research Centre for Physics, Budapest, Hungary
\item \Idef{org146}Yale University, New Haven, Connecticut, United States
\item \Idef{org147}Yonsei University, Seoul, Republic of Korea
\end{Authlist}
\endgroup
  
\end{document}